\documentclass[a4paper,11pt]{article}
\pdfoutput=1 
\usepackage{jheppub} 
\usepackage{amsmath,amsfonts,amssymb}
\usepackage{graphicx}
\usepackage{color}
\usepackage[dvipsnames,svgnames,x11names]{xcolor}
\usepackage{slashed} 
\usepackage{url}
\usepackage{subfigure}
\usepackage{multirow} 
\usepackage[utf8]{inputenc} 

\graphicspath{{fig/}} 

\def\TeV{\mathrm{TeV}} 
\def\GeV{\mathrm{GeV}} 

\makeatother

\allowdisplaybreaks 

\renewcommand\arraystretch{1.3} 

\title{Triplet-Quadruplet Dark Matter}

\preprint{UCI-HEP-TR-2015-25}

\author[a]{Tim M.P.~Tait}
\author[a,b,c]{and Zhao-Huan Yu}

\affiliation[a]{Department of Physics and Astronomy, University of California, \\4129 Frederick Reines Hall, Irvine, California 92697, U.S.A.}
\affiliation[b]{Key Laboratory of Particle Astrophysics,\\
Institute of High Energy Physics, Chinese Academy of Sciences, \\19B Yuquan Road, Beijing 100049, China}
\affiliation[c]{ARC Centre of Excellence for Particle Physics at the Terascale,\\
School of Physics, The University of Melbourne, \\Tin Alley, Melbourne, Victoria 3010, Australia}

\emailAdd{ttait@uci.edu}
\emailAdd{zhao-huan.yu@unimelb.edu.au}

\abstract{We explore a dark matter model extending the standard model particle content by one fermionic $SU(2)_L$ triplet and two 
fermionic $SU(2)_L$ quadruplets, leading to a minimal realistic UV-complete model of electroweakly interacting dark matter
which interacts with the Higgs doublet at tree level via two kinds of Yukawa couplings.
After electroweak symmetry-breaking, the physical spectrum of
the dark sector consists of three Majorana fermions, three singly charged fermions, and one doubly charged fermion,
with the lightest neutral fermion $\chi_1^0$ serving as a dark matter candidate.
A typical spectrum exhibits a large degree of degeneracy in mass between the neutral and charged fermions,
and we examine the one-loop corrections to the mass differences to ensure that the lightest particle is neutral.
We identify regions of parameter space for which the dark matter abundance is saturated for a standard cosmology,
including coannihilation channels, and find that this is typically achieved for 
$m_{\chi_1^0}\sim 2.4~\TeV$.  Constraints from
precision electroweak measurements, searches for dark matter scattering with nuclei,
and dark matter annihilation are important, but leave open a viable range for a thermal relic.}


\begin{document}

\maketitle
\flushbottom

\section{Introduction}
\label{sec:intro}

With the discovery of the $\sim 125~\GeV$ Higgs boson at the 
LHC~\cite{Aad:2012tfa,Chatrchyan:2012xdj}, the Standard Model (SM) of particle physics has been proven to be a self-consistent 
$SU(3)_C \times SU(2)_L \times U(1)_Y$ gauge theory describing the strong and electroweak interactions of three generation quarks 
and leptons.
However, the SM fails to describe astrophysical and cosmological 
observations, which are best explained by the existence of a massive neutral species of particle -- 
dark matter (DM)~\cite{Jungman:1995df,Bertone:2004pz,Feng:2010gw}.
While a variety of DM candidates are provided by extensions of the SM, among the most attractive are
weakly interacting massive particles (WIMPs), which have roughly weak interaction strength and 
masses of $\mathcal{O}(\GeV) - \mathcal{O}(\TeV)$.
If WIMPs were thermally produced in the early Universe, they could give a desired relic abundance consistent with observation.

WIMPs typically appear in popular extensions of the SM aimed at addressing its deficiencies,
such as e.g. supersymmetric~\cite{Goldberg:1983nd,Ellis:1983ew} and extra dimensional models~\cite{Servant:2002aq,Cheng:2002ej}.
However, the need for their existence is independent of deep theoretical questions and it behooves us to leave no stone unturned
in exploring the full range of possibilities.  It is further natural to explore dark sectors containing
$SU(2)_L$ multiplets, whose neutral components are natural DM candidates and whose interactions suggest the correct
relic density for weak scale masses.  
Within the broad class of such models, both theoretical considerations and experimental results (most importantly, 
the null results of searches for WIMP scattering with heavy nuclei) provide 
important constraints on the 
viable constructions.

In {\em minimal} dark matter~\cite{Cirelli:2005uq}, the dark sector consists
of a single scalar or fermion in a non-trivial $SU(2)_L$ representation.  For even-dimensional $SU(2)_L$ representations, non-zero hypercharge
is required to engineer an electrically neutral component, and typically results in a large coupling to the $Z$ boson, which is excluded
by direct searches for dark matter~\cite{Essig:2007az}.  Odd-dimensional $SU(2)_L$ representations have much weaker constraints, 
and lead to thermal relics for masses in the range of a few TeV.

If the dark sector consists of more than one $SU(2)_L$ representation, electroweak symmetry-breaking allows for mixing
between them, resulting in
a much richer theoretical landscape.  If the dark matter is a fermion, tree level renormalizable
couplings to the Standard Model Higgs are permitted provided there are $SU(2)_L$ representations differing in dimensionality
by one.  Such theories provide a theoretical laboratory to explore the possibility that the dark matter communicates to the SM
predominantly via exchange of the electroweak and Higgs bosons\footnote{In contract, scalar dark matter can always couple to
the Higgs via renormalizable quartic interactions.  We restrict our discussion to the fermionic case, leaving exploration of
scalar dark sectors for future work.}.  The minimal module consists of a single odd-dimensional $SU(2)_L$ representation
Weyl fermion together with a vector-like pair 
(such that anomalies cancel) of even-dimensional representations with an appropriate hypercharge.
Two such constructions which have been previously considered are
{\em singlet-doublet dark matter}~\cite{Mahbubani:2005pt,D'Eramo:2007ga,Enberg:2007rp,Cohen:2011ec,Calibbi:2015nha,Freitas:2015hsa} 
and {\em doublet-triplet} dark matter~\cite{Dedes:2014hga,Freitas:2015hsa}.  Both of these sets look (in the appropriate limit) like
subsets of the neutralino sector of the minimal supersymmetric standard model (MSSM), and share some of its phenomenology.

In this work we investigate a case which does not emerge simply as a limit of the MSSM, {\em triplet-quadruplet} dark matter,
consisting of one Weyl $SU(2)_L$ triplet with $Y=0$ and two Weyl quadruplets with $Y=\pm 1/2$.  
After electroweak symmetry-breaking, the mass eigenstates include three neutral Majorana fermions 
$\chi_i^0$, three singly charged fermions $\chi_i^\pm$, and one doubly charged fermion $\chi^{\pm\pm}$, leading to
unique features in the phenomenology.
After imposing a discrete $Z_2$ symmetry, and choosing the lightest neutral fermion $\chi_1^0$ to be lighter than its
charged siblings, we arrive at an exotic theory of dark matter whose interactions are mediated by the electroweak and
Higgs bosons.

As with the singlet-doublet and
doublet-triplet constructions, this theory is described by four parameters encapsulating two 
gauge-invariant mass terms ($m_T$ and $m_Q$) and two different Yukawa interactions coupling them to the
SM Higgs doublet ($y_1$ and $y_2$).  The limit $y_1 = y_2$ realizes an enhanced custodial global symmetry
resulting in $\chi_1^0$ decoupling (at tree level) from the $Z$ and Higgs bosons (provided $m_Q<m_T$),
greatly weakening the bounds from direct searches.  It further implies that $\chi_1^0$ is degenerate in
mass with one of the charged states (and sometimes $\chi^{\pm\pm}$) at tree level.  For small
deviations from this limit, the degeneracy is mildly lifted, requiring inclusion of the one-loop corrections
to reliably establish that the lightest dark sector fermion is neutral.

This paper is outlined as follows.
In Sec.~\ref{sec:model} we describe triplet-quadruplet dark matter in detail and establish notation.
In Sec.~\ref{sec:custo} we discuss the interesting features in the custodial symmetry limit.
In Sec.~\ref{sec:mass} we compute the corrections to the mass splittings at the one-loop level.
In Sec.~\ref{sec:phen} we identify the regions of parameter space resulting in the correct thermal relic abundance 
for a standard cosmology (including coannihilation channels)
as well as the constraints from the electroweak oblique parameters and from direct and indirect searches.
Sec.~\ref{sec:concl} contains our conclusions and further discussions.
Appendix~\ref{app:int} gives the explicit expressions for the interaction terms, while Appendix~\ref{app:self} lists the self-energy expressions which are used in the calculations of the mass corrections and electroweak oblique parameters.

\section{Triplet-Quadruplet Dark Matter}
\label{sec:model}

The triplet-quadruplet dark sector consists of colorless Weyl fermions $T$, $Q_1$, and $Q_2$ transforming under
$(SU(2)_L, U(1)_Y)$ as $(\textbf{3}, 0)$, $(\textbf{4},-1/2)$, and $(\textbf{4}, +1/2)$. We denote their components as:
\begin{eqnarray}
T = \left( {\begin{array}{c}
   {{T^ + }}  \\
   {{T^0}}  \\
   {{T^ - }}  \\
 \end{array} } \right),~~~~~~
{Q_1} = \left( {\begin{array}{c}
   {Q_1^ + }  \\
   {Q_1^0}  \\
   {Q_1^ - }  \\
   {Q_1^{ -  - }}  \\
 \end{array} } \right),~~~~~~
{Q_2} = \left( {\begin{array}{c}
   {Q_2^{ +  + }}  \\
   {Q_2^ + }  \\
   {Q_2^0}  \\
   {Q_2^ - }  \\
 \end{array} } \right).
\label{eq:vec}
\end{eqnarray}
The two quadruplets are assigned opposite hypercharges in order to cancel gauge anomalies.
Gauge-invariant kinetic and mass terms for the triplet and the quadruplets are given by
\begin{equation}
{\mathcal{L}_{\mathrm{T}}} = i{T^\dag }{{\bar \sigma }^\mu }{D_\mu }T - \frac{1}{2}({m_T}~TT + \mathrm{h.c.})
\label{eq:Lag_T}
\end{equation}
and
\begin{equation}
{\mathcal{L}_{\mathrm{Q}}} = iQ_1^\dag {{\bar \sigma }^\mu }{D_\mu }{Q_1} + iQ_2^\dag {{\bar \sigma }^\mu }{D_\mu }{Q_2} 
- ({m_Q}~{Q_1}{Q_2} + \mathrm{h.c.}),
\label{eq:Lag_Q}
\end{equation}
which specify their interactions with electroweak gauge bosons.
They also couple to the SM Higgs doublet $H$ through Yukawa interactions
\begin{equation}
{\mathcal{L}_{{\mathrm{HTQ}}}} = {y_1}~{\varepsilon _{jl}}({Q_1})_i^{jk}T_k^i{H^l} 
- {y_2}~({Q_2})_i^{jk}T_k^iH_j^\dag  + \mathrm{h.c.}\,,
\label{eq:Lag_HTQ}
\end{equation}
where we use the tensor notation (see e.g. Ref.~\cite{Georgi:1982jb}) 
to write down the triplet and quadruplets with $SU(2)_L$ $\mathbf{2}$ (upper) and $\mathbf{\bar{2}}$ 
(lower) indices explicitly indicated.
We further assume there is a $Z_2$ symmetry under which dark sector fermions are odd while SM particles are even 
to forbid renormalizable operators $TLH$ and nonrenormalizable operators such as 
$TeHH$, $Q_1L^\dag HH^\dag$, and $Q_2LHH^\dag$ (where $L$ is a lepton doublet and $e$ is a charged lepton singlet),
which would lead the lightest dark sector fermion to decay.

In decomposing the $SU(2)$ components, a traceless tensor $\mathcal{T}_j^i$ in the $\mathbf{3}$ 
representation is constructed from a $\mathbf{2}$, $u^i$, and a $\mathbf{\bar{2}}$, $v_i$, as
\begin{equation}
\mathcal{T}_j^i = {u^i}{v_j} - \frac{1}{2}\delta _j^i{u^k}{v_k},
\label{eq:tensor_T}
\end{equation}
whereas a $\mathbf{4}$, $\mathcal{Q}_k^{ij}$, is constructed via
\begin{equation}
\mathcal{Q}_k^{ij} = \frac{1}{2}\left( {\mathcal{T}_k^i{u^j} + \mathcal{T}_k^j{u^i} - \frac{1}{3}\delta _k^i\mathcal{T}_l^j{u^l} - \frac{1}{3}\delta _k^j\mathcal{T}_l^i{u^l}} \right),
\label{eq:tensor_Q}
\end{equation}
which is symmetric in the upper indices $i$ and $j$, and satisfies $\sum_k \mathcal{Q}_k^{kj} = \sum_k \mathcal{Q}_k^{ik} = 0$.
Taking into account the normalization of the Lagrangians \eqref{eq:Lag_T} and \eqref{eq:Lag_Q},
we can identify the components of $T$, $Q_1$, and $Q_2$ in the vector notation \eqref{eq:vec} with those in the tensor notation via:
\begin{eqnarray}
&& {T^ + } = T_2^1,~
{T^0} = \sqrt 2 T_1^1 =  - \sqrt 2 T_2^2,~
{T^ - } = T_1^2;
\\
&& Q_1^ +  = ({Q_1})_2^{11},~
Q_1^0 = \sqrt 3 ({Q_1})_1^{11} =  - \sqrt 3 ({Q_1})_2^{12} =  - \sqrt 3 ({Q_1})_2^{21},
\\
&& Q_1^ -  = \sqrt 3 ({Q_1})_2^{22} =  - \sqrt 3 ({Q_1})_1^{12} =  - \sqrt 3 ({Q_1})_1^{21},~
Q_1^{ -  - } = ({Q_1})_1^{22};
\\
&& Q_2^{ +  + } = ({Q_2})_2^{11},~
Q_2^ +  = \sqrt 3 ({Q_2})_1^{11} =  - \sqrt 3 ({Q_2})_2^{12} =  - \sqrt 3 ({Q_2})_2^{21},
\\
&& Q_2^0 = \sqrt 3 ({Q_2})_2^{22} =  - \sqrt 3 ({Q_2})_1^{12} =  - \sqrt 3 ({Q_2})_1^{21},~
Q_2^ -  = ({Q_2})_1^{22}.
\end{eqnarray}
Thus, the mass terms decompose into
\begin{equation}
 - \frac{1}{2}{m_T}TT \equiv  - \frac{1}{2}{m_T}T_i^jT_j^i =  - {m_T}{T^ - }{T^ + } - \frac{1}{2}{m_T}{T^0}{T^0}
\end{equation}
and
\begin{equation}
 - {m_Q}{Q_1}{Q_2} \equiv  
 - {m_Q}{\varepsilon _{il}}({Q_1})_k^{ij}({Q_2})_j^{lk} =  - {m_Q}(Q_1^{ -  - }Q_2^{ +  + } - Q_1^ - Q_2^ +  + Q_1^0Q_2^0 - Q_1^ + Q_2^ - ).~~~
\end{equation}
The explicit form of the Higgs doublet is
\begin{equation}
{H^i} = \left( {\begin{array}{*{20}{c}}
   {{H^ + }}  \\
   {{H^0}}  \\
 \end{array} } \right),~
 H_i^\dag  = ( {{H^ - }}, {{H^{0*}}}),
\end{equation}
leading to
\begin{equation}
{H^i(x)} = \frac{1}{{\sqrt 2 }}\left( {\begin{array}{*{20}{c}}
   0  \\
   {v + h(x)}  \\
 \end{array} } \right)
\end{equation}
after electroweak symmetry-breaking in the unitary gauge. Then
\begin{eqnarray}
{\mathcal{L}_{{\mathrm{HTQ}}}} & \rightarrow &
{y_1}(v + h)\left( {\frac{1}{{\sqrt 6 }}Q_1^ - {T^ + } - \frac{1}{{\sqrt 3 }}Q_1^0{T^0} - \frac{1}{{\sqrt 2 }}Q_1^ + {T^ - }} \right)
\nonumber\\
&& + {y_2}(v + h)\left( {\frac{1}{{\sqrt 3 }}Q_2^0{T^0} + \frac{1}{{\sqrt 6 }}Q_2^ + {T^ - } - \frac{1}{{\sqrt 2 }}Q_2^ - {T^ + }} \right).
\end{eqnarray}

The complete model-dependence is specified by the four parameters,
\begin{equation}
\left\{ m_T,~~m_Q,~~y_1,~~y_2 \right\}.
\end{equation}
By choosing appropriate field redefinitions, $m_T$, $y_1$, and $y_2$ can be made to be real, such that 
the phase of $m_Q$ is the only source of $CP$ violation in the dark sector.  However, here
we do not consider $CP$ violation effects and take all of them to be real.
Moreover, taking $m_T \to  - m_T$, the transformation $m_Q \to  - m_Q$ or $y_2 \to - y_2$ each yields 
the same Lagrangian up to field redefinitions.
Therefore, we consider $m_T$ and $m_Q$ both positive without
loss of generality.

After electroweak symmetry breaking, the full set of mass terms can be written
\begin{eqnarray}
{\mathcal{L}_{{\mathrm{mass}}}} &=&  - {m_Q}Q_1^{ -  - }Q_2^{ +  + } 
- \frac{1}{2} \left({T^0},Q_1^0,Q_2^0 \right){\mathcal{M}_N}\left( {\begin{array}{*{20}{c}}
   {{T^0}}  \\
   {Q_1^0}  \\
   {Q_2^0}  \\
 \end{array} } \right)
 - \left({T^ - },Q_1^ - ,Q_2^ - \right){\mathcal{M}_C}\left( {\begin{array}{*{20}{c}}
   {{T^ + }}  \\
   {Q_1^ + }  \\
   {Q_2^ + }  \\
 \end{array} } \right) + \mathrm{h.c.}
\nonumber\\
 &=&  - {m_Q}~{\chi ^{ -  - }}{\chi ^{ +  + }} - \frac{1}{2}\sum\limits_{i = 1}^3 {{m_{\chi _i^0}}~\chi _i^0\chi _i^0}
 - \sum\limits_{i = 1}^3 {{m_{\chi _i^ \pm }}~\chi _i^ - \chi _i^ + }  + \mathrm{h.c.}\,,
\label{eq:lag_mass}
\end{eqnarray}
where $\chi^{--}\equiv Q_1^{--}$ and $\chi^{++}\equiv Q_2^{++}$.
The mass matrices for the neutral and charged fermions are given by
\begin{eqnarray}
\hspace*{-0.75cm}
{\mathcal{M}_N} = \left( {\renewcommand\arraystretch{2.2}\begin{array}{*{20}{c}}
   {{m_T}} & {~~\dfrac{1}{{\sqrt 3 }}{y_1}v~~} & { - \dfrac{1}{{\sqrt 3 }}{y_2}v}  \\
   {\dfrac{1}{{\sqrt 3 }}{y_1}v} & 0 & {{m_Q}}  \\
   { - \dfrac{1}{{\sqrt 3 }}{y_2}v} & {{m_Q}} & 0  \\
 \end{array} } \right),~
{\mathcal{M}_C} = \left( {\renewcommand\arraystretch{2.2}\begin{array}{*{20}{c}}
   {{m_T}} & {~~\dfrac{1}{{\sqrt 2 }}{y_1}v~~} & { - \dfrac{1}{{\sqrt 6 }}{y_2}v}  \\
   { - \dfrac{1}{{\sqrt 6 }}{y_1}v} & 0 & { - {m_Q}}  \\
   {\dfrac{1}{{\sqrt 2 }}{y_2}v} & { - {m_Q}} & 0  \\
 \end{array} } \right).\quad
\end{eqnarray}
They are diagonalized by three unitary matrices, $\mathcal{N}$, $\mathcal{C}_L$, and $\mathcal{C}_R$:
\begin{eqnarray}
{\mathcal{N}^{\mathrm{T}}}{\mathcal{M}_N}\mathcal{N} &=& {\tilde {\mathcal{M}}_N} = {\mathrm{diag}}({m_{\chi _1^0}},{m_{\chi _2^0}},{m_{\chi _3^0}}),
\\
\mathcal{C}_R^{\mathrm{T}}{\mathcal{M}_C}{\mathcal{C}_L} &=& {\tilde{\mathcal{M}}_C} = {\mathrm{diag}}({m_{\chi _1^ \pm }},{m_{\chi _2^ \pm }},{m_{\chi _3^ \pm }}),
\end{eqnarray}
with the gauge eigenstates related to the mass eigenstates by
\begin{equation}
\left( {\begin{array}{*{20}{c}}
   {{T^0}}  \\
   {Q_1^0}  \\
   {Q_2^0}  \\
 \end{array} } \right) = \mathcal{N}\left( {\begin{array}{*{20}{c}}
   {\chi _1^0}  \\
   {\chi _2^0}  \\
   {\chi _3^0}  \\
 \end{array} } \right),~
\left( {\begin{array}{*{20}{c}}
   {{T^ + }}  \\
   {Q_1^ + }  \\
   {Q_2^ + }  \\
 \end{array} } \right) = {\mathcal{C}_L}\left( {\begin{array}{*{20}{c}}
   {\chi _1^ + }  \\
   {\chi _2^ + }  \\
   {\chi _3^ + }  \\
 \end{array} } \right),~
\left( {\begin{array}{*{20}{c}}
   {{T^ - }}  \\
   {Q_1^ - }  \\
   {Q_2^ - }  \\
 \end{array} } \right) = {\mathcal{C}_R}\left( {\begin{array}{*{20}{c}}
   {\chi _1^ - }  \\
   {\chi _2^ - }  \\
   {\chi _3^ - }  \\
 \end{array} } \right).
\label{eq:state_mix}
\end{equation}
Therefore, the dark sector fermions consist of three Majorana fermions $\chi_i^0$, 
three singly charged fermions $\chi_i^\pm$, and one doubly charged fermion $\chi^{\pm\pm}$.
Here we denote the particles in order of mass, i.e., $m_{\chi_1^0} \leq m_{\chi_2^0} \leq m_{\chi_3^0}$ 
and $m_{\chi_1^\pm} \leq m_{\chi_2^\pm} \leq m_{\chi_3^\pm}$.
The lightest new particle is stable as a result of the imposed $Z_2$ symmetry. 
Consequently, we identify parameters such that $\chi_1^0$ is 
lighter than $\chi_1^\pm$ and $\chi^{\pm\pm}$, in order for $\chi_1^0$ to effectively play the role of dark matter.

We can construct 4-component fermionic fields from the Weyl fields:
\begin{equation}
X_i^0 = \left( {\begin{array}{*{20}{c}}
   {\chi _{iL}^0}  \\
   {{{(\chi _{iR}^0)}^\dag }}  \\
 \end{array} } \right),~
X_i^ +  = \left( {\begin{array}{*{20}{c}}
   {\chi _{iL}^ + }  \\
   {{{(\chi _{iR}^ - )}^\dag }}  \\
 \end{array} } \right),~
{X^{ +  + }} = \left( {\begin{array}{*{20}{c}}
   {\chi _L^{ +  + }}  \\
   {{{(\chi _R^{ -  - })}^\dag }}  \\
 \end{array} } \right),
\end{equation}
where
\begin{equation}
\chi _L^0 = \chi _R^0 = {(\chi _1^0,\chi _2^0,\chi _3^0)^{\mathrm{T}}},~
\chi _L^ +  = {(\chi _1^ + ,\chi _2^ + ,\chi _3^ + )^{\mathrm{T}}},~
\chi _R^ -  = {(\chi _1^ - ,\chi _2^ - ,\chi _3^ - )^{\mathrm{T}}},
\end{equation}
\begin{equation}
\chi _L^{ +  + } = {\chi ^{ +  + }},~
\chi _R^{ -  - } = {\chi ^{ -  - }}.
\end{equation}
And the mass basis is defined such that they have diagonal mass terms:
\begin{equation}
{\mathcal{L}_{{\mathrm{mass}}}} =  - {m_Q}~{{\bar X}^{ +  + }}{X^{ +  + }} 
- \frac{1}{2}\sum\limits_{i = 1}^3 {{m_{\chi _i^0}}~\bar X_i^0X_i^0}  
- \sum\limits_{i = 1}^3 {{m_{\chi _i^ \pm }}~\bar X_i^ + X_i^ + }.
\end{equation}

\section{Custodial Symmetry}
\label{sec:custo}

If $y_1$ is equal to $y_2$, there exists a global custodial $SU(2)_R$ global symmetry, as is well known in the SM Higgs sector.
Under this symmetry the triplet is an $SU(2)_R$ singlet, while the quadruplets and the Higgs field are both $SU(2)_R$ doublets:
\begin{equation}
({{\mathbf{Q}}_A})_k^{ij} = \left( {\begin{array}{*{20}{c}}
   {({Q_1})_k^{ij}}  \\
   {({Q_2})_k^{ij}}  \\
 \end{array} } \right),~
{({{\mathbf{H}}_A})_i} = \left( {\begin{array}{*{20}{c}}
   {H_i^\dag }  \\
   {{H_i}}  \\
 \end{array} } \right),
\end{equation}
where ${H_i} \equiv {\varepsilon _{ij}}{H^j}$ and  $A$ is an $SU(2)_R$ index.
$\mathcal{L}_\mathrm{Q}$ and $\mathcal{L}_\mathrm{HTQ}$ can be expressed in an $SU(2)_L\times SU(2)_R$ invariant form:
\begin{eqnarray}
{\mathcal{L}_{\mathrm{Q}}} + {\mathcal{L}_{{\mathrm{HTQ}}}}
&=& i({\mathbf{Q}^{\dag A}})_{ij}^k{{\bar \sigma }^\mu }{D_\mu }({\mathbf{Q}_A})_k^{ij} 
- \frac{1}{2}{m_Q}\:\left[{\varepsilon ^{AB}}{\varepsilon _{il}}({\mathbf{Q}_A})_k^{ij}({\mathbf{Q}_B})_j^{lk} + \mathrm{h.c.}\right]
\nonumber\\*
&& + \left[y~{\varepsilon ^{AB}}({\mathbf{Q}_A})_i^{jk}T_k^i{({\mathbf{H}_B})_j} + \mathrm{h.c.}\right],
\end{eqnarray}
where $y=y_1=y_2$.
This symmetry is also found in the singlet-doublet model~\cite{D'Eramo:2007ga,Enberg:2007rp,Cohen:2011ec} 
and the doublet-triplet model~\cite{Dedes:2014hga}.  Though
broken by the $U(1)_Y$ gauge symmetry, nonetheless it dictates some tree level relations with important implications.  We describe
the cases $m_Q < m_T$ and $m_Q > m_T$ separately below.

\subsection{$m_Q < m_T$}

\begin{figure}[t]
\centering
\subfigure[$m_Q < m_T$ case.\label{fig:custo:mass_spec:a}]
{\includegraphics[width=.49\textwidth]{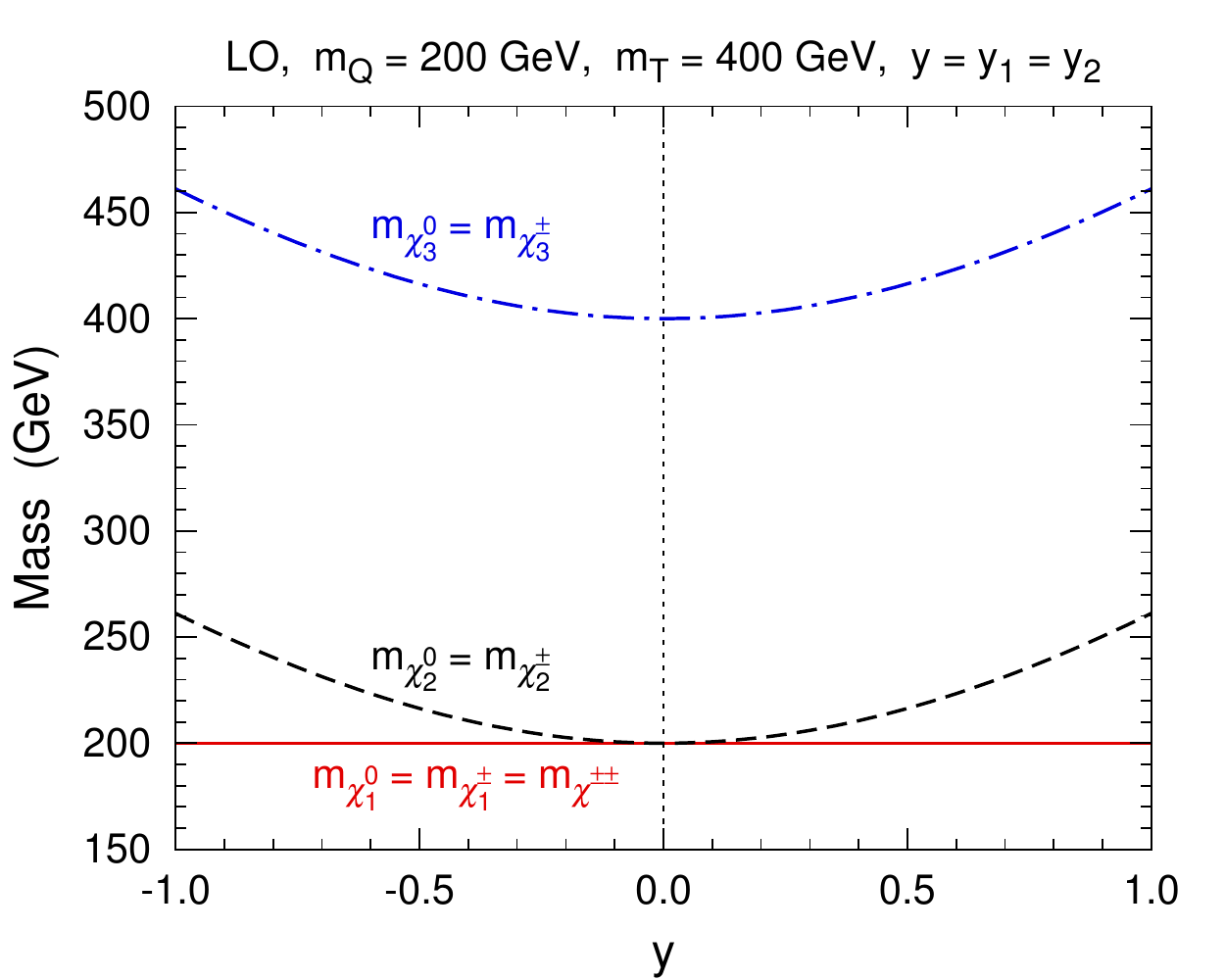}}
\subfigure[$m_T < m_Q$ case.\label{fig:custo:mass_spec:b}]
{\includegraphics[width=.49\textwidth]{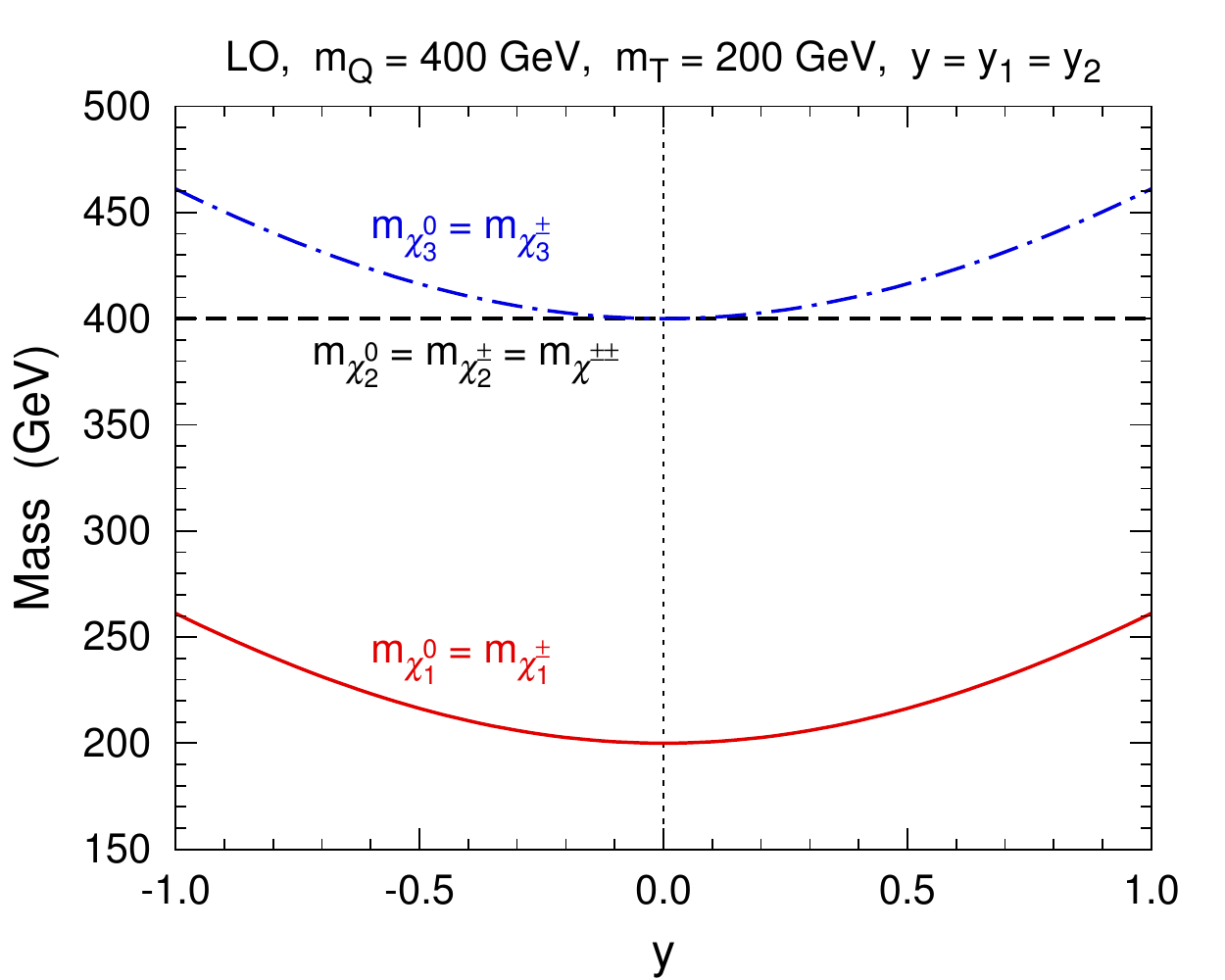}}
\caption{Fermion masses as functions of $y$ in the custodial symmetry limit at LO.
The left (right) panel corresponds to $m_Q = 200~(400)~\GeV$ and $m_T = 400~(200)~\GeV$.}
\label{fig:custo:mass_spec}
\end{figure}

If $m_Q < m_T$, the leading order (LO) dark sector fermion masses can be derived to be:
\begin{eqnarray}
{m_{\chi _1^0}^\mathrm{LO}} &=& {m_{\chi _1^ \pm }^\mathrm{LO}} = {m_{{\chi ^{ \pm  \pm }}}^\mathrm{LO}} = {m_Q},
\\
{m_{\chi _2^0}^\mathrm{LO}} &=& {m_{\chi _2^ \pm }^\mathrm{LO}} = \frac{1}{2}\left[ {\sqrt {8{y^2}{v^2}/3 + {{({m_Q} + {m_T})}^2}}  + {m_Q} - {m_T}} \right],
\\
{m_{\chi _3^0}^\mathrm{LO}} &=& {m_{\chi _3^ \pm }^\mathrm{LO}} = \frac{1}{2}\left[ {\sqrt {8{y^2}{v^2}/3 + {{({m_Q} + {m_T})}^2}}  - {m_Q} + {m_T}} \right],
\end{eqnarray}
while the mixing matrices take the form
\begin{equation}
\mathcal{N} = \left( {\begin{array}{*{20}{c}}
   0 & {\frac{{ai}}{b}} & { - \frac{{\sqrt 2 }}{b}}  \\
   {\frac{1}{{\sqrt 2 }}} & { - \frac{i}{b}} & { - \frac{a}{{\sqrt 2 b}}}  \\
   {\frac{1}{{\sqrt 2 }}} & {\frac{i}{b}} & {\frac{a}{{\sqrt 2 b}}}  \\
 \end{array} } \right),~
{\mathcal{C}_L} = \left( {\begin{array}{*{20}{c}}
   0 & {\frac{a}{b}} & { - \frac{{\sqrt 2 i}}{b}}  \\
   {\frac{i}{2}} & { - \frac{{\sqrt 6 }}{{2b}}} & { - \frac{{\sqrt 3 ai}}{{2b}}}  \\
   {\frac{{\sqrt 3 i}}{2}} & {\frac{{\sqrt 2 }}{{2b}}} & {\frac{{ai}}{{2b}}}  \\
 \end{array} } \right),~
{\mathcal{C}_R} = \left( {\begin{array}{*{20}{c}}
   0 & { - \frac{a}{b}} & {\frac{{\sqrt 2 i}}{b}}  \\
   {\frac{{\sqrt 3 i}}{2}} & { - \frac{{\sqrt 2 }}{{2b}}} & { - \frac{{ai}}{{2b}}}  \\
   {\frac{i}{2}} & {\frac{{\sqrt 6 }}{{2b}}} & {\frac{{\sqrt 3 ai}}{{2b}}}  \\
 \end{array} } \right),
\label{eq:mixing_custo}
\end{equation}
where
\begin{equation}
a \equiv \frac{{\sqrt {8{y^2}{v^2}/3 + {{({m_Q} + {m_T})}^2}}  - {m_Q} - {m_T}}}{{2yv/\sqrt 3 }}
~~\text{and}~~
b \equiv \sqrt {2 + {a^2}}.
\end{equation}
Thus each of the neutral fermions is degenerate in mass with a singly charged fermion, and the lightest one is also degenerate with the doubly charged fermion, which always has a mass of $m_Q$.
In Fig.~\ref{fig:custo:mass_spec:a}, we show the mass spectrum for $m_Q = 200~\GeV$ and $m_T = 400~\GeV$.
If $y=0$, the quadruplets would not mix with the triplet, and we would have $m_{\chi _1^0}^{{\mathrm{LO}}} = m_{\chi _1^ \pm }^{{\mathrm{LO}}} = m_{\chi _2^0}^{{\mathrm{LO}}} = m_{\chi _2^ \pm }^{{\mathrm{LO}}}= m_{{\chi ^{ \pm  \pm }}}^{{\mathrm{LO}}} = {m_Q}$ and $m_{\chi _3^0}^{{\mathrm{LO}}} = m_{\chi _3^ \pm }^{{\mathrm{LO}}} = m_T$.
As $|y|$ increases, $\chi_2^0$, $\chi_3^0$, $\chi_2^\pm$, and $\chi_3^\pm$ become heavier.
At loop level the custodial symmetry realizes that it is broken by $U(1)_Y$, and 
corrections from the loops of electroweak bosons lift the degeneracies ~\cite{Feng:1999fu,Cirelli:2005uq,Hill:2011be}.
We examine the next-to-leading (NLO) corrections to the masses in detail in Sec.~\ref{sec:mass}.

In general, the $\chi_1^0$ couplings to the Higgs boson and to the $Z$ boson are proportional to $({y_1}{\mathcal{N}_{21}} - {y_2}{\mathcal{N}_{31}})\mathcal{N}_{11}$ and $(|\mathcal{N}_{31}|^2 - |\mathcal{N}_{21}|^2)$, respectively.
In the custodial symmetry limit, the interaction properties of $\chi_1^0$ are quite special.
From the explicit expression of $\mathcal{N}$ in Eq.~\eqref{eq:mixing_custo}, we can find that there is no triplet component in $\chi _1^0$ and $\mathcal{N}_{21}=\mathcal{N}_{31}=1/\sqrt{2}$, i.e., $\chi _1^0 = (Q_1^0 + Q_2^0)/\sqrt 2$.
Therefore, the $\chi_1^0$ coupling to the Higgs boson vanishes because this coupling exists only when the $T^0$ component is involved.
Moreover, there is no $\chi_1^0$ coupling to the $Z$ boson, since $Q_1^0$ and $Q_2^0$ have opposite hypercharges and opposite eigenvalues of the third $SU(2)_L$ generator.
As a result, $\chi_1^0$ cannot interact with nuclei at tree level and generically escapes from direct detection bounds.

\subsection{$m_Q > m_T$}

If $m_Q > m_T$ and $|yv| < \sqrt {3{m_Q}({m_Q} - {m_T})}$, the fermion masses are
\begin{eqnarray}
m_{\chi _1^0}^{{\mathrm{LO}}} &=& m_{\chi _1^ \pm }^{{\mathrm{LO}}} = \frac{1}{2}\left[ {\sqrt {8{y^2}{v^2}/3 + {{({m_Q} + {m_T})}^2}}  - {m_Q} + {m_T}} \right],
\\
m_{\chi _2^0}^{{\mathrm{LO}}} &=& m_{\chi _2^ \pm }^{{\mathrm{LO}}} = m_{{\chi ^{ \pm  \pm }}}^{{\mathrm{LO}}} = {m_Q},
\\
m_{\chi _3^0}^{{\mathrm{LO}}} &=& m_{\chi _3^ \pm }^{{\mathrm{LO}}} = \frac{1}{2}\left[ {\sqrt {8{y^2}{v^2}/3 + {{({m_Q} + {m_T})}^2}}  + {m_Q} - {m_T}} \right],
\end{eqnarray}
and $\chi_1^0$ is a mixture of $T^0$, $Q_1^0$, and $Q_2^0$:
\begin{equation}
\chi _1^0 =  - \frac{i}{b}(a{T^0} - Q_1^0 + Q_2^0).
\end{equation}
In this case, the coupling to the Higgs boson does not vanish, that with 
the $Z$ boson still vanishes because $|\mathcal{N}_{21}|^2 = |\mathcal{N}_{31}|^2=1/b^2$.
Consequently, $\chi_1^0$ can interact with nuclei through the Higgs exchange at tree level.
Fig.~\ref{fig:custo:mass_spec:b} shows the mass spectrum for $m_Q = 400~\GeV$ and $m_T = 200~\GeV$.

If $m_Q > m_T$ and $|yv| > \sqrt {3{m_Q}({m_Q} - {m_T})}$,
we have
\begin{equation*}
m_Q < \frac{1}{2}\left[ {\sqrt {8{y^2}{v^2}/3 + {{({m_Q} + {m_T})}^2}}  - {m_Q} + {m_T}} \right],
\end{equation*}
and hence $m_{\chi_1^0}=m_Q$ and $\chi_1^0 = (Q_1^0 + Q_2^0)/\sqrt 2$, whose interactions are similar to the case of $m_Q < m_T$
described above.

\section{One Loop Mass Corrections}
\label{sec:mass}

In this section, we calculate the dark fermion mass corrections at NLO,
determining the parameter space for which $\chi_1^0$ is lighter than $\chi_1^\pm$ and $\chi^{\pm\pm}$.

For mixed fermionic fields $X_i$ (either $X_i^0$ or $X_i^+$),
renormalized one-particle irreducible two-point functions can be written down as~\cite{Baro:2009gn,Denner:1991kt}
\begin{eqnarray}
{\hat\Sigma _{{X_i}{X_j}}}(q) &=& (\slashed q - {m_{{\chi _i}}}){\delta _{ij}} + {\Sigma _{{X_i}{X_j}}}(q) - \delta {{ \mathcal{\tilde M}}_{ij}}{P_L} - \delta \mathcal{\tilde M}_{ji}^*{P_R}
\nonumber\\
&& + \frac{1}{2}(\slashed q - {m_{{\chi _i}}})(\delta Z_{ij}^L{P_L} + \delta Z_{ij}^{R*}{P_R}) + \frac{1}{2}(\delta Z_{ji}^{L*}{P_R} + \delta Z_{ji}^R{P_L})(\slashed q - {m_{{\chi _j}}}),
\end{eqnarray}
where ${P_L} \equiv \frac{1}{2}(1 - {\gamma _5})$ and ${P_R} \equiv \frac{1}{2}(1 + {\gamma _5})$ are chiral projectors and
$\delta\mathcal{\tilde M}_{ij}$ are mass renormalization constants defined by 
${{\mathcal{\tilde M}}_{ij,0}} = {{\mathcal{\tilde M}}_{ij}} + \delta {{\mathcal{\tilde M}}_{ij}}$, 
where the subscript $0$ denotes a bare quantity and the diagonalized mass matrix 
$\mathcal{\tilde M}$ stands for either $\mathcal{\tilde M}_N$ or $\mathcal{\tilde M}_C$.
The wave function renormalization constants $\delta Z_{ij}^L$ and $\delta Z_{ij}^R$ are defined as
${X_{i,0}} = {X_i} + \frac{1}{2}(\delta Z_{ij}^L{P_L} + \delta Z_{ij}^{R*}{P_R}){X_j}$.
The self-energy ${\Sigma _{{X_i}{X_j}}}(q)$ can be decomposed into Lorentz structures:
\begin{equation}
{\Sigma _{{X_i}{X_j}}}(q) = {P_L}\Sigma _{{X_i}{X_j}}^{{\mathrm{LS}}}({q^2}) + {P_R}\Sigma _{{X_i}{X_j}}^{{\mathrm{RS}}}({q^2}) 
+ \slashed q{P_L}\Sigma _{{X_i}{X_j}}^{{\mathrm{LV}}}({q^2}) + \slashed q{P_R}\Sigma _{{X_i}{X_j}}^{{\mathrm{RV}}}({q^2}),
\end{equation}
and Hermiticity relates these functions:
\begin{equation}
\Sigma _{{X_i}{X_j}}^{{\mathrm{RS}}}({q^2}) = \Sigma _{{X_j}{X_i}}^{{\mathrm{LS}}*}({q^2}),~~~~~ 
\Sigma _{{X_i}{X_j}}^{{\mathrm{LV}}}({q^2}) = \Sigma _{{X_j}{X_i}}^{\mathrm{LV}*}({q^2}),~~~~~
\Sigma _{{X_i}{X_j}}^{{\mathrm{RV}}}({q^2}) = \Sigma _{{X_j}{X_i}}^{{\mathrm{RV}}*}({q^2}).
\end{equation}
There are additional constraints for Majorana fields $X_i^0$:
\begin{equation}
\Sigma _{X_i^0X_j^0}^{{\mathrm{LS}}}({q^2}) = \Sigma _{X_j^0X_i^0}^{{\mathrm{LS}}}({q^2}),~~~~~ 
\Sigma _{X_i^0X_j^0}^{{\mathrm{RS}}}({q^2}) = \Sigma _{X_j^0X_i^0}^{{\mathrm{RS}}}({q^2}),~~~~~ 
\Sigma _{X_i^0X_j^0}^{{\mathrm{LV}}}({q^2}) = \Sigma _{X_j^0X_i^0}^{{\mathrm{RV}}}({q^2}),
\end{equation}
which we utilize as a cross-check on our calculations.

On-shell, there should be no mixing between states in the mass basis. 
Using the definition of the pole mass in the on-shell scheme leads to the renormalization condition:
\begin{equation}
\widetilde{\operatorname{Re} }~{\hat\Sigma _{{X_i}{X_j}}}(q){u_{{X_j}}}(q) = 0 ~~\text{for}~~ {q^2} = m_{{\chi _j}}^2,
\end{equation}
where $\widetilde{\operatorname{Re}}$ takes the real parts of the loop integrals in self-energies but leaves the couplings intact.
This condition fixes the mass renormalization constants to
\begin{equation}
\delta {{\mathcal{\tilde M}}_{ij}} = \frac{1}{2}\widetilde{\operatorname{Re} }[\Sigma _{{X_i}{X_j}}^{{\mathrm{LS}}}(m_{{\chi _i}}^2) 
+ \Sigma _{{X_i}{X_j}}^{{\mathrm{LS}}}(m_{{\chi _j}}^2) 
+ {m_{{\chi _i}}}\Sigma _{{X_i}{X_j}}^{{\mathrm{LV}}}(m_{{\chi _i}}^2) 
+ {m_{{\chi _j}}}\Sigma _{{X_i}{X_j}}^{{\mathrm{RV}}}(m_{{\chi _j}}^2)].
\label{eq:mass_ren_const}
\end{equation}

As in Refs.~\cite{Fritzsche:2002bi,Baro:2009gn} for the renormalization of neutralinos and charginos, 
we introduce renormalization constants $\delta\mathcal{M}_N$ and $\delta\mathcal{M}_C$ to shift the mass 
matrices $\mathcal{M}_N$ and $\mathcal{M}_C$, but the mixing matrices $\mathcal{N}$, $\mathcal{C}_L$, and $\mathcal{C}_R$ remain
the same at NLO as at LO.
Therefore, we have
\begin{equation}
{(\delta {\mathcal{M}_N})_{ij}} = {({\mathcal{N}^*}\delta {{\mathcal{\tilde M}}_N}{\mathcal{N}^\dag })_{ij}} 
= \mathcal{N}_{ik}^*\mathcal{N}_{jl}^*{(\delta {{\mathcal{\tilde M}}_N})_{kl}},
\label{eq:dM_N:renorm}
\end{equation}
and
\begin{equation}
{(\delta {{\mathcal{\tilde M}}_C})_{ij}} = {(\mathcal{C}_R^{\mathrm{T}}\delta {\mathcal{M}_C}{\mathcal{C}_L})_{ij}} = {({\mathcal{C}_R})_{ki}}{({\mathcal{C}_L})_{lj}}{(\delta {\mathcal{M}_C})_{kl}}.
\label{eq:dM_C:renorm}
\end{equation}
Furthermore, we choose to renormalize the Majorana fermion masses on-shell, i.e.,
\begin{equation}
m_{\chi _i^0}^{{\mathrm{NLO}}} = {m_{\chi _i^0}},
\end{equation}
and compute the relative shifts in the masses of $\chi_i^\pm$ and $\chi^{\pm\pm}$.
In this scheme Eq.~\eqref{eq:dM_N:renorm} provides the NLO shifts in the parameters $m_T$, $m_Q$, $y_1$, and $y_2$:
\begin{eqnarray}
\delta {m_T} &=& \mathcal{N}_{1k}^*\mathcal{N}_{1l}^*{(\delta {{\mathcal{\tilde M}}_N})_{kl}},~~~
\delta {m_Q} = \mathcal{N}_{2k}^*\mathcal{N}_{3l}^*{(\delta {{\mathcal{\tilde M}}_N})_{kl}},
\\
v\delta {y_1} &=& \sqrt 3 \mathcal{N}_{1k}^*\mathcal{N}_{2l}^*{(\delta {{\mathcal{\tilde M}}_N})_{kl}},~~~
v\delta {y_2} =  - \sqrt 3 \mathcal{N}_{1k}^*\mathcal{N}_{3l}^*{(\delta {{\mathcal{\tilde M}}_N})_{kl}},
\end{eqnarray}
where $\delta {{\mathcal{\tilde M}}_N}$ is given by Eq.~\eqref{eq:mass_ren_const}:
\begin{equation}
{(\delta {{\mathcal{\tilde M}}_N})_{ij}} = \frac{1}{2}\widetilde{\operatorname{Re} }
\left[\Sigma _{X_i^0X_j^0}^{{\mathrm{LS}}}(m_{\chi _i^0}^2) 
+ \Sigma _{X_i^0X_j^0}^{{\mathrm{LS}}}(m_{\chi _j^0}^2) 
+ {m_{\chi _i^0}}\Sigma _{X_i^0X_j^0}^{{\mathrm{LV}}}(m_{\chi _i^0}^2) 
+ {m_{\chi _j^0}}\Sigma _{X_i^0X_j^0}^{{\mathrm{RV}}}(m_{\chi _j^0}^2)\right].
\end{equation}
The shifts on these parameters shift $\mathcal{\tilde M}_C$ through Eq.~\eqref{eq:dM_C:renorm}.
As a result, the physical masses of $\chi_i^\pm$ at NLO are given by
\begin{equation}
m_{\chi _i^ \pm }^{{\mathrm{NLO}}} = {m_{\chi _i^ \pm }} + {(\delta {{\mathcal{\tilde M}}_C})_{ii}} - \frac{1}{2}\widetilde{\operatorname{Re} }\{ 2\Sigma _{X_i^ + X_i^ + }^{{\mathrm{LS}}}(m_{\chi _i^ \pm }^2) + {m_{\chi _i^ \pm }}[\Sigma _{X_i^ + X_i^ + }^{{\mathrm{LV}}}(m_{\chi _i^ \pm }^2) + \Sigma _{X_i^ + X_i^ + }^{{\mathrm{RV}}}(m_{\chi _i^ \pm }^2)]\},
\end{equation}
where
\begin{eqnarray}
{(\delta {{\mathcal{\tilde M}}_C})_{ii}} &=& \sum\limits_{jk} {({\mathcal{C}_R})_{ji}}{({\mathcal{C}_L})_{ki}}
{(\delta {\mathcal{M}_C})_{jk}} \nonumber\\
&=& - \left[{({\mathcal{C}_R})_{2i}}{({\mathcal{C}_L})_{3i}} + {({\mathcal{C}_R})_{3i}}{({\mathcal{C}_L})_{2i}}]\delta {m_Q} 
+ \frac{1}{{\sqrt 6 }}v\delta {y_1}[\sqrt 3 {({\mathcal{C}_R})_{1i}}{({\mathcal{C}_L})_{2i}} 
- {({\mathcal{C}_R})_{2i}}{({\mathcal{C}_L})_{1i}}\right] \nonumber\\
&& + {({\mathcal{C}_R})_{1i}}{({\mathcal{C}_L})_{1i}}\delta {m_T} 
+ \frac{1}{{\sqrt 6 }}v\delta {y_2}
\left[\sqrt 3 {({\mathcal{C}_R})_{3i}}{({\mathcal{C}_L})_{1i}} - {({\mathcal{C}_R})_{1i}}{({\mathcal{C}_L})_{3i}}\right].
\end{eqnarray}
The physical mass of $\chi^{\pm\pm}$ is affected by the shift in $m_Q$:
\begin{equation}
m_{{\chi ^{ \pm  \pm }}}^{{\mathrm{NLO}}} = {m_Q} + \delta {m_Q} 
- \frac{1}{2}\widetilde{\operatorname{Re}} 
\left\{ 2 \Sigma _{{X^{ +  + }X^{ +  + }}}^{{\mathrm{LS}}}(m_{{\chi ^{ \pm  \pm }}}^2) 
+ {m_Q}[\Sigma _{{X^{ +  + }X^{ +  + }}}^{{\mathrm{LV}}}(m_{{\chi ^{ \pm  \pm }}}^2) 
+ \Sigma _{{X^{ +  + }X^{ +  + }}}^{{\mathrm{RV}}}(m_{{\chi ^{ \pm  \pm }}}^2)]\right\}.
\end{equation}

\begin{figure}[!t]
\centering
\subfigure[$m_Q < m_T$ case.]
{\includegraphics[width=.49\textwidth]{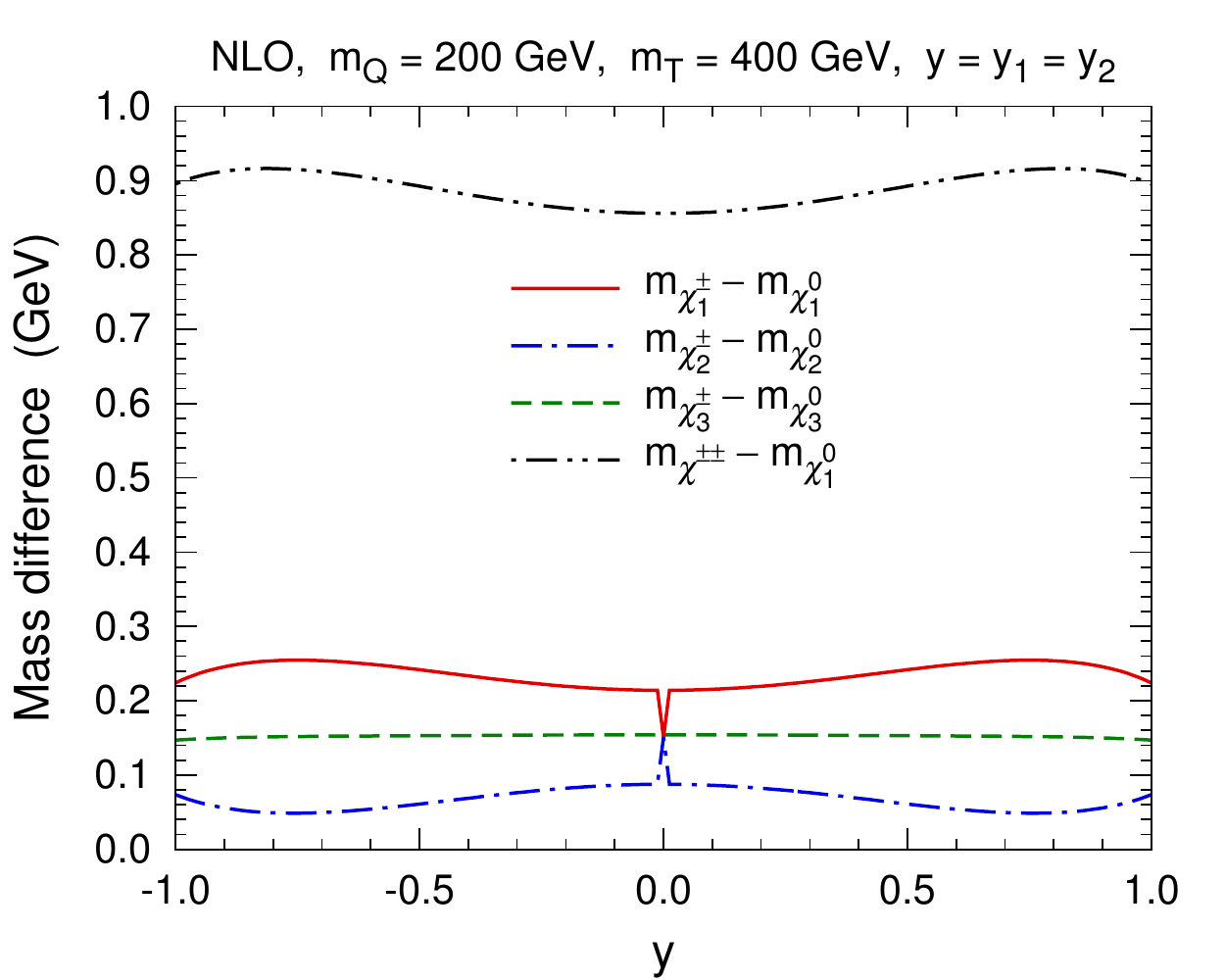}}
\subfigure[$m_T < m_Q$ case.]
{\includegraphics[width=.49\textwidth]{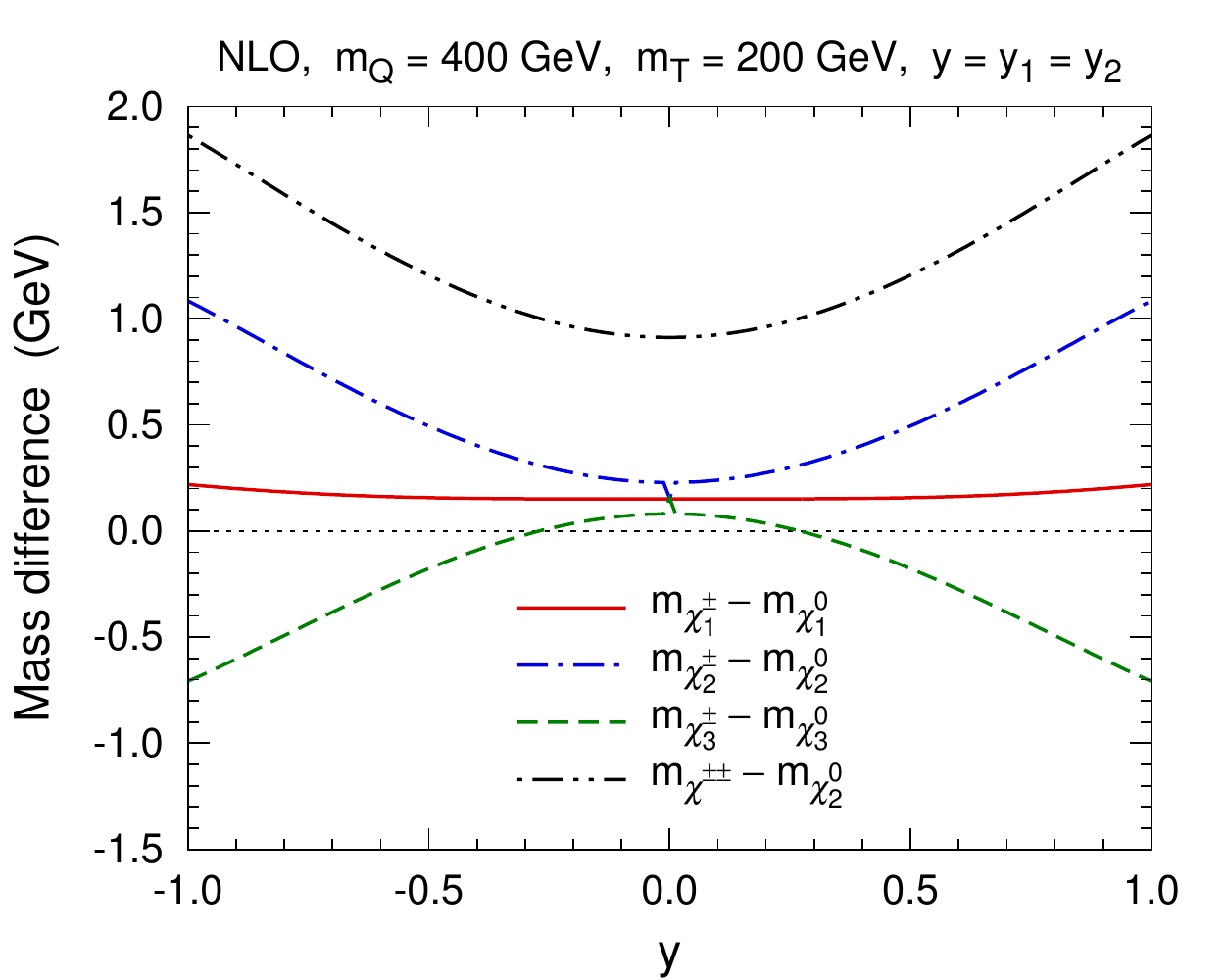}}
\caption{NLO mass differences between charged and neutral fermions in the custodial symmetry limit $y=y_1=y_2$.
The left (right) panel corresponds to $m_Q = 200~(400)~\GeV$ and $m_T = 400~(200)~\GeV$.}
\label{fig:custo:mass_diff}
\end{figure}

Explicit expressions for the self-energies of dark sector fermions at NLO can be found in Appendix~\ref{app:self}.
We evaluate the mass corrections numerically  with \texttt{LoopTools}~\cite{Hahn:1998yk}.
In the custodial symmetry limit $y=y_1=y_2$, the mass differences between charged and neutral 
fermions at NLO are presented in Fig.~\ref{fig:custo:mass_diff}.
$m_{\chi_1}^\pm - m_{\chi_1}^0$, $m_{\chi_2}^\pm - m_{\chi_2}^0$, and $m_{\chi_3}^\pm - m_{\chi_3}^0$ 
are degenerate for $y=0$, where the triplet has no mixing with the quadruplets, and the mass splitting is solely induced by the irreducible $\mathcal{O}(100)~\mathrm{MeV}$ contribution from the electroweak gauge interaction at one loop~\cite{Cirelli:2005uq}.
This degeneracy lifts for $y\ne 0$. 
When $m_Q=200~\GeV$ and $m_T=400~\GeV$, the charged fermions are always heavier than their 
corresponding neutral fermions for $|y|\leq 1$.
When $m_Q=400~\GeV$ and $m_T=200~\GeV$, $\chi_3^\pm$ becomes lighter than 
$\chi_3^0$ for $0.25\lesssim|y|\leq 1$.
In both cases, $\chi_1^0$ is always the lightest dark sector fermion as required for a DM candidate.

\begin{figure}[!t]
\centering
\subfigure[$m_Q < m_T$ case.]
{\includegraphics[width=.49\textwidth]{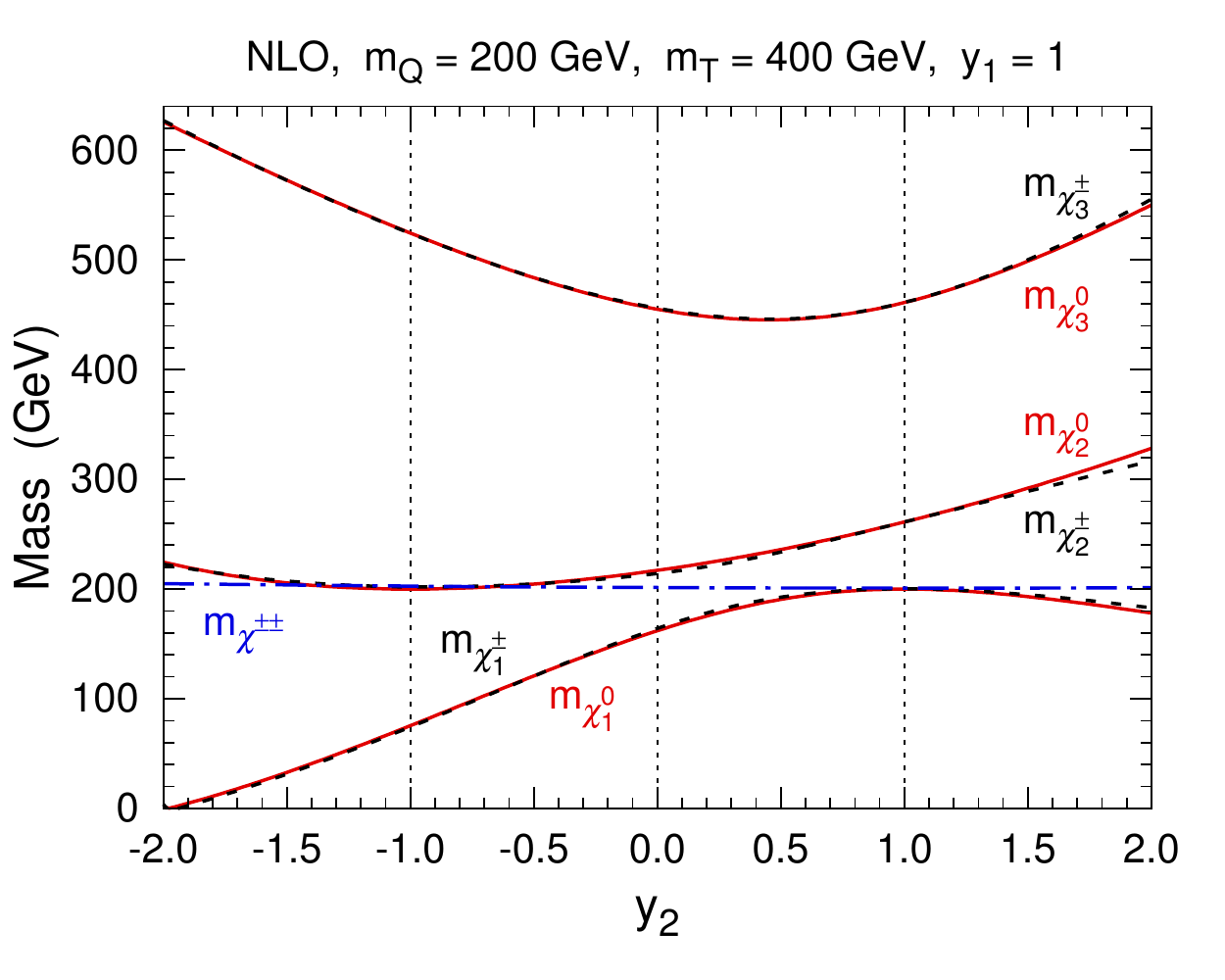}}
\subfigure[$m_T < m_Q$ case.]
{\includegraphics[width=.49\textwidth]{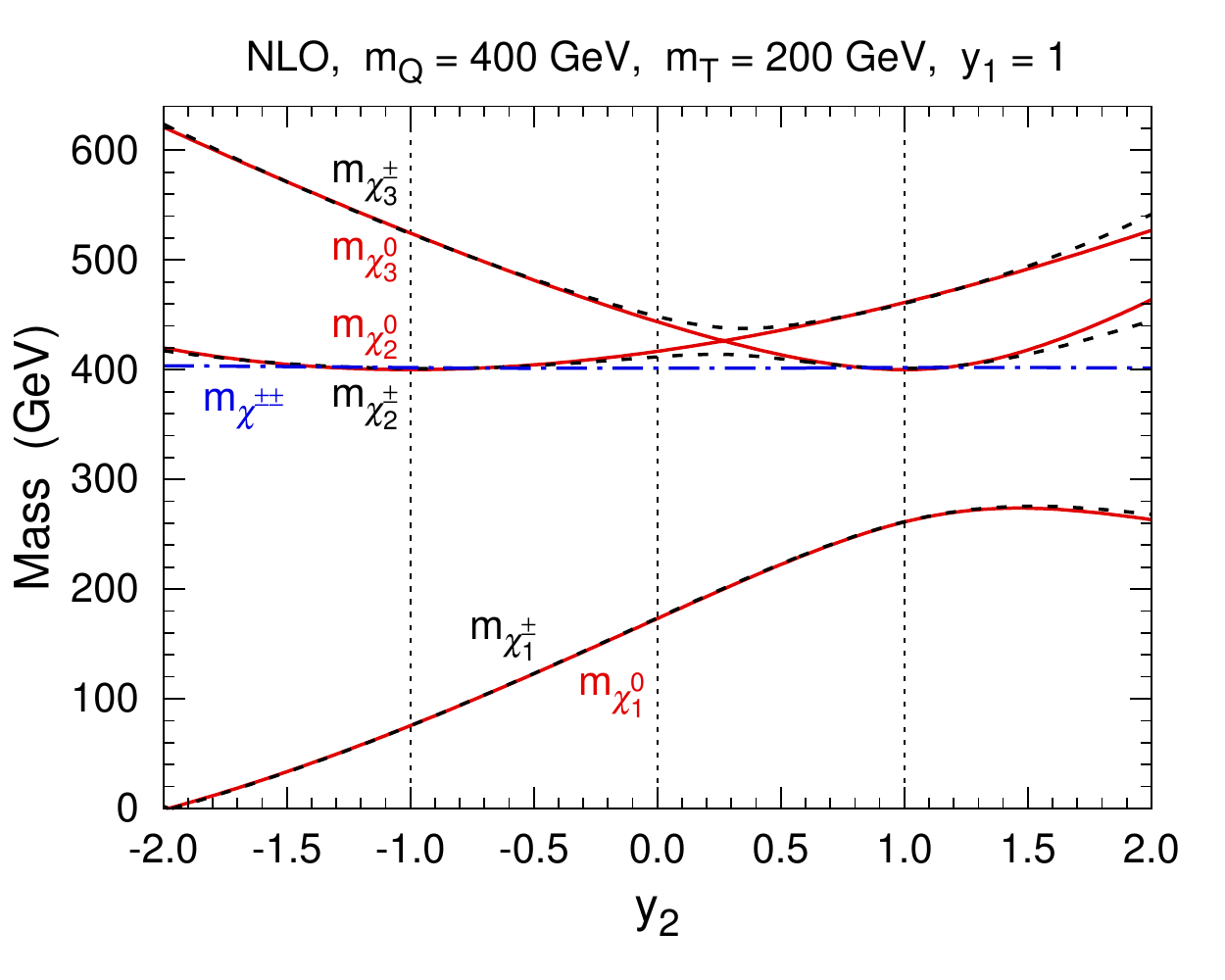}}
\caption{NLO fermion masses as functions of $y_2$ for $y_1=1$.
In the left (right) panel, $m_Q = 200~(400)~\GeV$ and $m_T = 400~(200)~\GeV$.
The red solid lines correspond to the neutral fermions, while the black dashed and blue dot-dashed lines correspond to the singly and doubly charged fermions, respectively.}
\label{fig:mass_spec}
\end{figure}

\begin{figure}[!htbp]
\centering
\subfigure[$m_Q < m_T$ case.]
{\includegraphics[width=.49\textwidth]{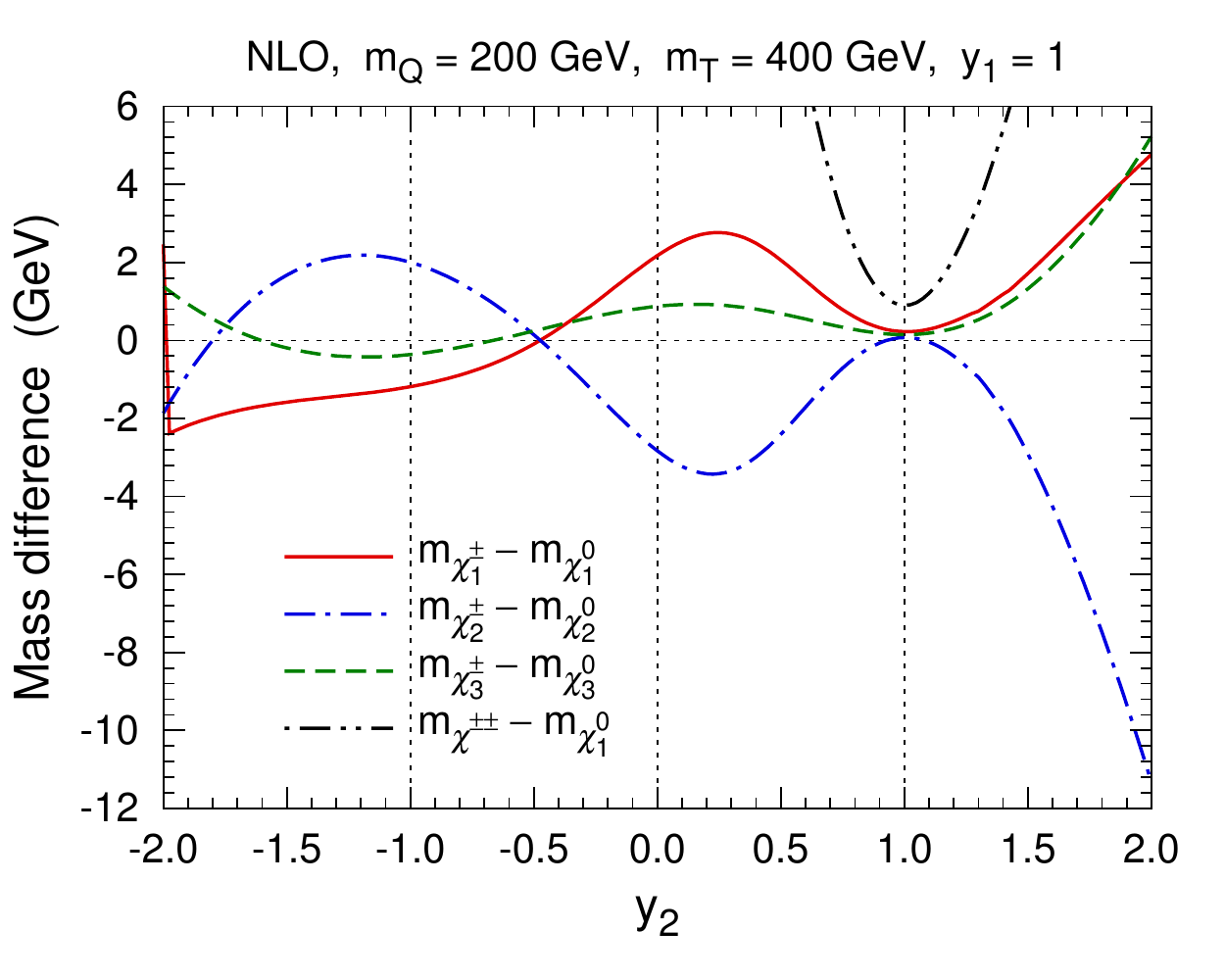}}
\subfigure[$m_T < m_Q$ case.]
{\includegraphics[width=.49\textwidth]{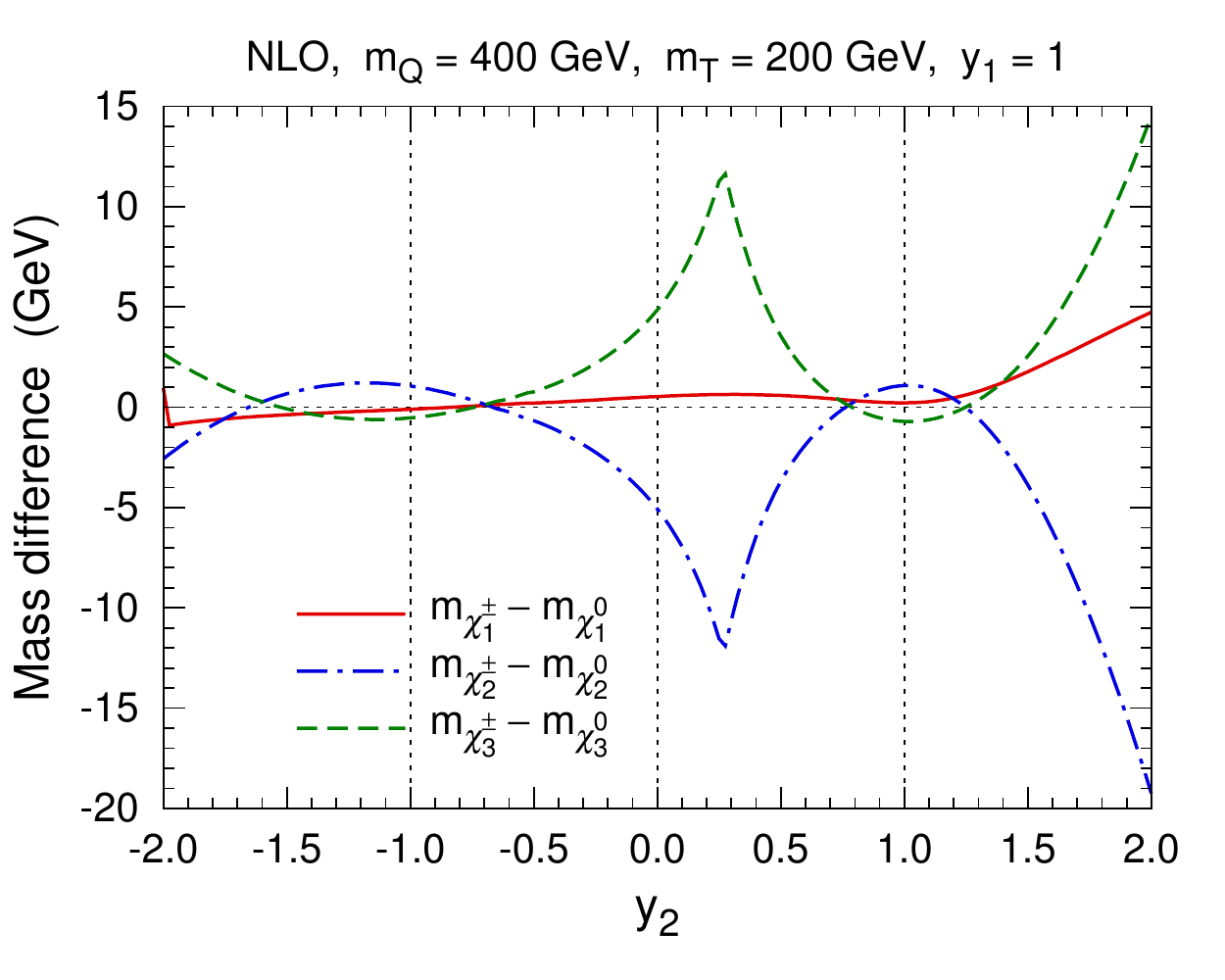}}
\caption{Mass differences at NLO between charged and neutral fermions as functions of $y_2$ for $y_1=1$.
In the left (right) panel, $m_Q = 200~(400)~\GeV$ and $m_T = 400~(200)~\GeV$.}
\label{fig:mass_diff}
\end{figure}

Moving beyond the custodial symmetry limit, in 
Fig.~\ref{fig:mass_spec}, we fix $m_Q$, $m_T$, and $y_1=1$, and  plot the fermion masses as functions of $y_2$.
We find that a value of $y_2$ unequal to $y_1$ tends to drive $\chi_1^0$ lighter, especially when the 
sign of $y_2$ is opposite to $y_1$.
The charged fermions remain rather degenerate with the corresponding neutral fermions.
In Fig.~\ref{fig:mass_diff}, we present the corresponding mass differences, which change sign frequently as $y_2$ varies.
For $-1.95 \lesssim y_2 \lesssim -0.5$ ($-1.95 \lesssim y_2 \lesssim -0.85$) in the $m_Q < m_T$ ($m_T < m_Q$) case, $\chi_1^0$ becomes lighter than $\chi_1^\pm$ and fails to describe viable DM.

\section{Constraints and Relic Density}
\label{sec:phen}

\begin{figure}[!t]
\centering
\subfigure[$y_1 = y_2 = 0.5$.\label{fig:limits:mQmT:a}]
{\includegraphics[width=.49\textwidth]{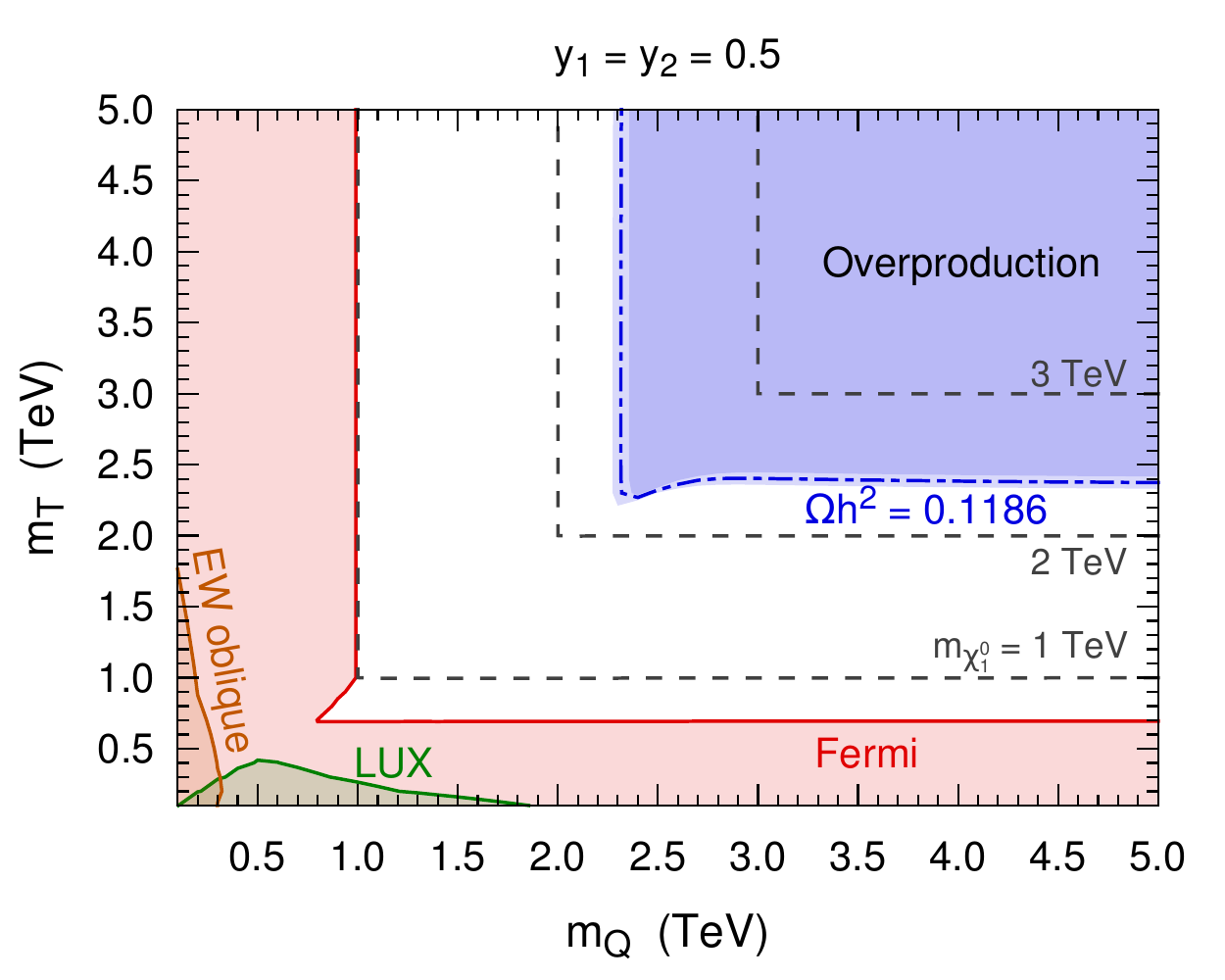}}
\subfigure[$y_1 = 0.5$, $y_2 = 1$.\label{fig:limits:mQmT:b}]
{\includegraphics[width=.49\textwidth]{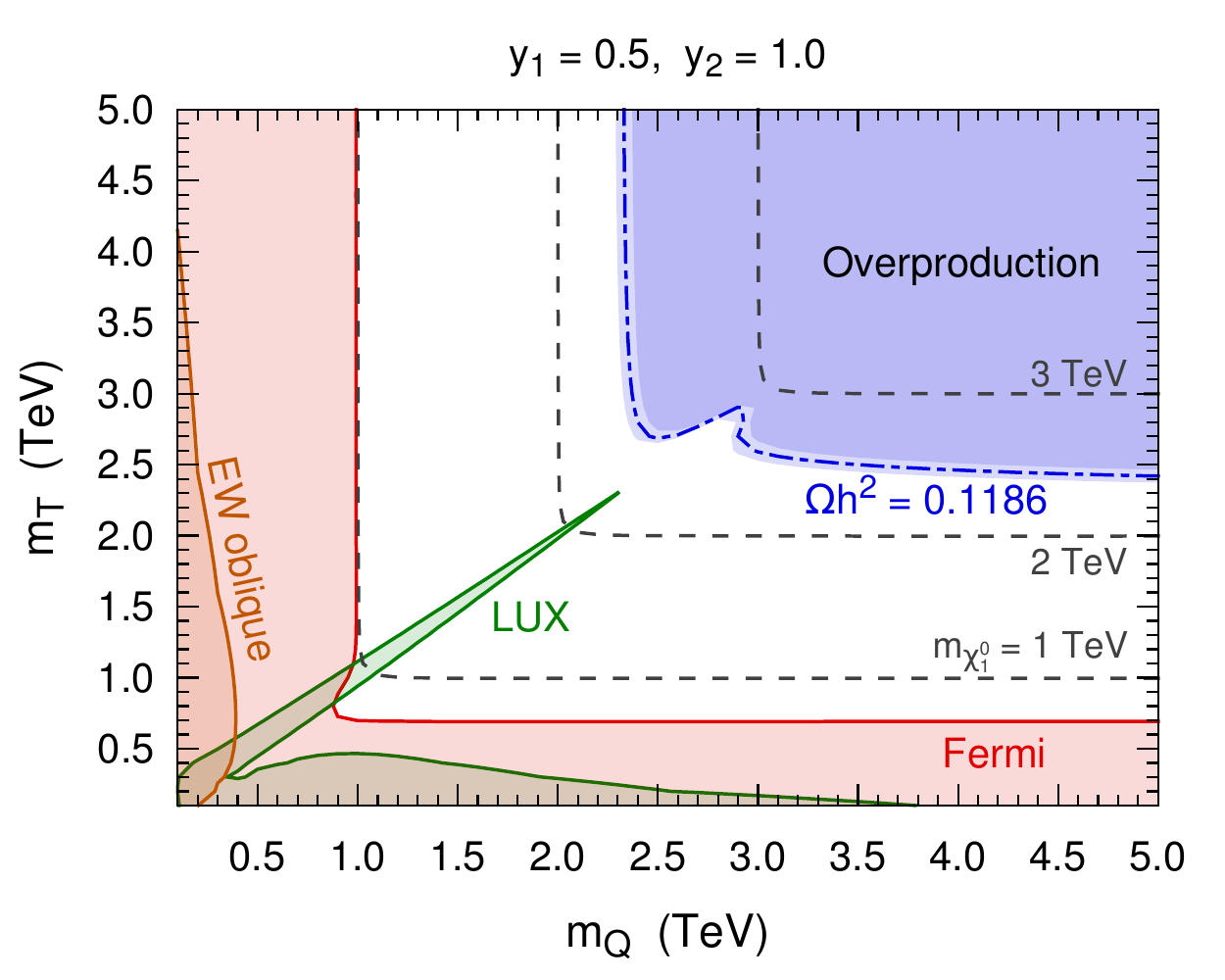}}
\subfigure[$y_1 = 0.5$, $y_2 = -0.5$.\label{fig:limits:mQmT:c}]
{\includegraphics[width=.49\textwidth]{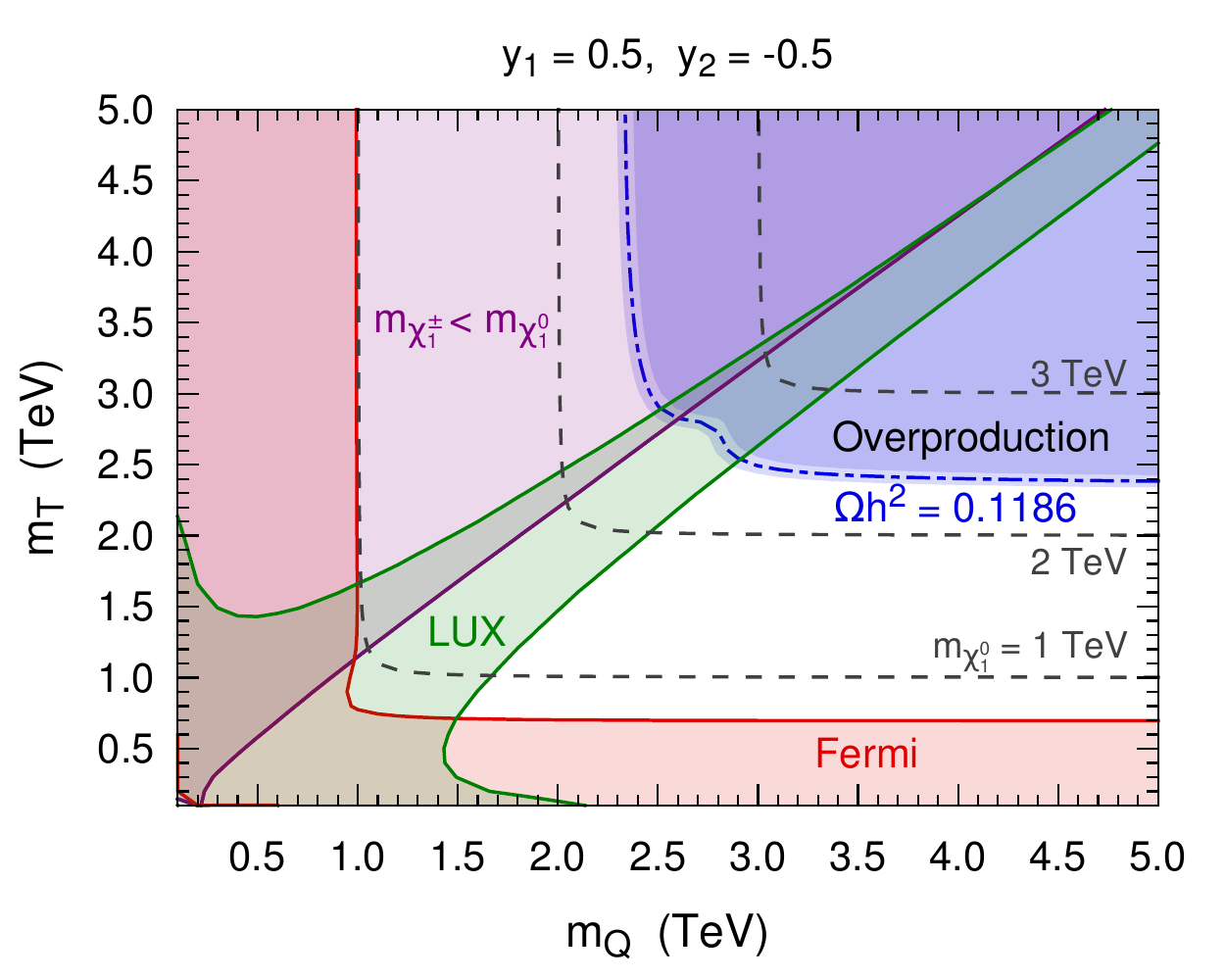}}
\subfigure[$y_1 = 0.5$, $y_2 = -1$.\label{fig:limits:mQmT:d}]
{\includegraphics[width=.49\textwidth]{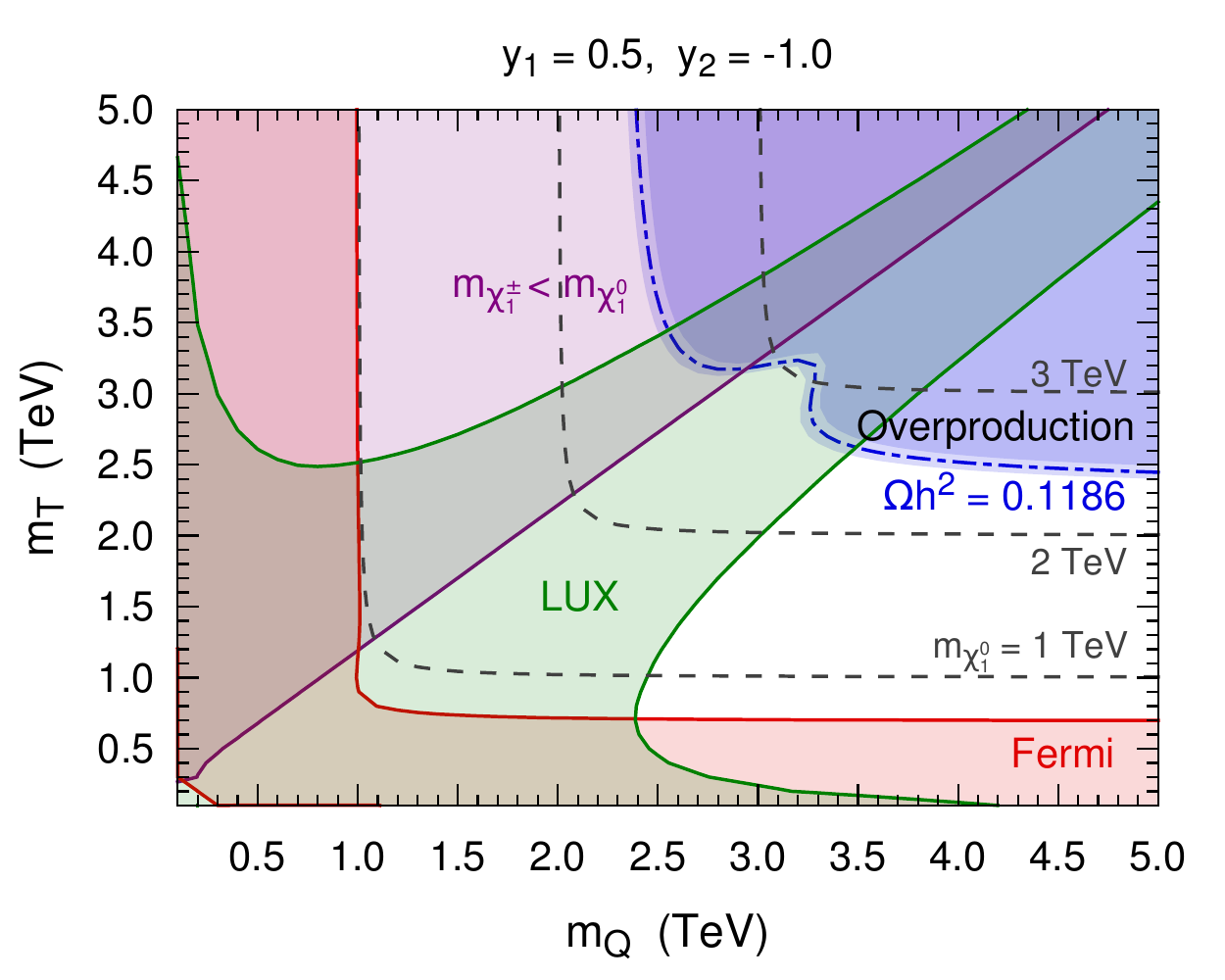}}
\caption{Constraints in the $m_Q$-$m_T$ plane for four sets of fixed $y_1$ and $y_2$.
The dot-dashed lines correspond to the mean value of the observed DM relic abundance~\cite{Ade:2015xua}, while the light blue bands 
denote its $2\sigma$ range and the dark blue regions indicate DM overproduction in the early Universe.
The violet, orange, green, and red regions are excluded by the condition $m_{\chi_1^\pm} < m_{\chi_1^0}$, 
electroweak oblique parameters~\cite{Baak:2014ora}, the LUX direct detection experiment~\cite{Akerib:2013tjd}, 
and the Fermi-LAT gamma-ray observations on dwarf galaxies~\cite{Ackermann:2015zua}, respectively.
The gray dashed lines indicate contours of fixed $m_{\chi_1^0}$.}
\label{fig:limits:mQmT}
\end{figure}

In this section, we investigate the constraints on the parameter space 
from electroweak precision measurements, direct and indirect searches,
and identify regions where the observed DM relic abundance is obtained for a standard cosmology.
We discuss each of these regions in greater detail below, but begin with a summary presented in 
Fig.~\ref{fig:limits:mQmT} in the $m_Q$-$m_T$ plane with the values of $y_1$ and $y_2$ fixed for four cases: 
(a) $y_1 = y_2 = 0.5$ (custodial symmetry limit); (b) $y_1 = 0.5$ and $y_2 = 1$; 
(c) $y_1 = 0.5$ and $y_2 = -0.5$; (d) $y_1 = 0.5$ and $y_2 = -1$.
The dashed lines in the plots denote the contours for $m_{\chi_1^0} = 1$, $2$, and $3~\TeV$.
When $y_1 v$ and $y_2 v$ are much smaller than $m_Q$ and $m_T$, $\chi_1^0$ is mainly constituted from the lighter multiplet.
Thus we find that $m_{\chi_1^0} \simeq m_Q$ for $m_Q<m_T$ and $m_{\chi_1^0} \simeq m_T$ for $m_T<m_Q$ in Fig.~\ref{fig:limits:mQmT}.
As we have seen in Fig.~\ref{fig:mass_diff}, when $y_2$ has a sign  opposite to $y_1$, $\chi_1^\pm$ may be lighter than $\chi_1^0$.
Therefore, in the cases (c) and (d) the condition $m_{\chi_1^\pm} < m_{\chi_1^0}$ (which implies that $\chi_1^0$ 
is not stable) excludes large portions of the parameter space, particularly 
when $m_Q<m_T$, as shown by the violet regions in Figs.~\ref{fig:limits:mQmT:c} and \ref{fig:limits:mQmT:d}.

\subsection{Relic Abundance}

To begin with, we identify the regions in which the dark matter abundance saturates observations for a standard cosmology.  As we
have seen,  $\chi_1^0$ is always nearly degenerate in mass with $\chi_1^\pm$.
Furthermore, for $m_Q<m_T$, we may have $m_{\chi^{\pm\pm}}\simeq m_{\chi_1^0}$, as well as $m_{\chi_2^0} \simeq m_{\chi_2^\pm} \simeq m_{\chi_1^0}$ when $m_Q \gg |y_{1,2}v|$.
These fermions, with close masses and comparable interaction strengths, tend to decouple 
at the same time, with coannihilation processes playing a significant role in their final abundances.
Since after freeze-out they decay into $\chi_1^0$, we compute their
combined relic abundance using the technology of Ref.~\cite{Griest:1990kh}.
We implement the triplet-quadruplet model in \texttt{Feynrules~2}~\cite{Alloul:2013bka}, and compute the relic density with
\texttt{MadDM}~\cite{Backovic:2013dpa} (based on \texttt{MadGraph~5}~\cite{Alwall:2014hca}).

In Fig.~\ref{fig:limits:mQmT}, the parameter space
consistent with the DM abundance measured by the Planck experiment, $\Omega h^2 = 0.1186\pm 0.020$~\cite{Ade:2015xua},
is plotted as the dot-dashed blue lines, with the $2\sigma$ region around it denoted by the light blue shading.
As is typical for an electroweakly-interacting WIMP, the observed DM abundance is realized for $m_{\chi_1^0} \sim 2.4~\TeV$.
When $\chi_1^0$ is heavier, there is effectively overproduction of DM in the early Universe, as shown by 
darker blue shaded regions in Fig.~\ref{fig:limits:mQmT}.  Regions with lighter masses and underproduction of dark matter are left unshaded.

\subsection{Precision Electroweak Constraints}

The dark fermions contribute at the one loop level to precision electroweak processes.  Since there are no direct coupling to the SM
fermions, these take the form of corrections to the electroweak boson propagators, and are encapsulated in the 
oblique parameters $S$, $T$, and $U$~\cite{Peskin:1990zt,Peskin:1991sw},
\begin{eqnarray}
S&\equiv&\frac{16\pi c_W^2s_W^2}{e^2}\left[\Pi'_{ZZ}(0)-
\frac{c_W^2-s_W^2}{c_Ws_W} \Pi'_{ZA}(0)-\Pi'_{AA}(0)\right],\\
T&\equiv&\frac{4\pi}{e^2}\left[\frac{\Pi_{WW}(0)}{m_W^2}-\frac{\Pi_{ZZ}(0)}{m_Z^2}\right],\\
U&\equiv&\frac{16\pi s_W^2}{e^2}\left[\Pi'_{WW}(0)-c_W^2\Pi'_{ZZ}(0)
-2c_W s_W \Pi'_{ZA}(0)
-s_W^2 \Pi'_{AA}(0)\right],
\end{eqnarray}
where $s_W\equiv\sin\theta_W$, $c_W\equiv\cos\theta_W$ with $\theta_W$ denoting the Weinberg angle.
$\Pi_{IJ}(p^2)$ is the $g_{\mu\nu}$ coefficient for the vacuum polarization amplitude of gauge bosons $I$ and $J$,
which can be divided as $i\Pi _{IJ}^{\mu \nu }({p^2}) = i{g^{\mu \nu }}{\Pi _{IJ}}({p^2}) + ({p^\mu }{p^\nu }~{\mathrm{terms}})$, and
$\Pi'_{IJ}(0)\equiv\partial\Pi_{IJ}(p^2)/\partial(p^2)|_{p^2=0}$. 

The contributions to $\Pi_{ZZ}(p^2)$, $\Pi_{WW}(p^2)$, $\Pi_{AA}(p^2)$, and $\Pi_{ZA}(p^2)$ 
from dark sector fermions are given in Appendix~\ref{app:self}.  In the
custodial symmetry limit, $T$ and $U$ remain zero, while $S$ is positive and increases as $|y|$ increases.  Outside of the
custodial limit, all are typically nonzero, with $U$ typically much smaller than $S$ and $T$, as is expected given the fact that
it corresponds to a higher dimensional operator.

A global fit to current measurements of precision data by the Gfitter Group yields~\cite{Baak:2014ora}
\begin{equation}
S = 0.05 \pm 0.11,~ T = 0.09 \pm 0.13,~ U = 0.01 \pm 0.11,
\end{equation}
with correlation coefficients,
\begin{equation}
{\rho _{ST}} =  + 0.90,~ {\rho _{SU}} =  - 0.59,~ {\rho _{TU}} =  - 0.83.
\end{equation}
These results exclude the orange regions in Figs.~\ref{fig:limits:mQmT:a} and \ref{fig:limits:mQmT:b} at the 95\% CL.
In the custodial symmetry limit $y_1 = y_2 = 0.5$, a region limited by $m_Q\lesssim 300~\GeV$ and $m_T\lesssim1.8~\TeV$ is excluded.
For $y_1 = 0.5$ and $y_2 = 1$, a region limited by $m_Q\lesssim 400~\GeV$ and $m_T\lesssim 4.1~\TeV$ is excluded.

\subsection{Scattering with Heavy Nuclei}

Spin-independent scattering with heavy nuclei is mediated at tree level by the exchange of a Higgs or $Z$ boson.
The coupling strength of $\chi^0_1$ to Higgs (see Appendix~\ref{app:int}) is:
\begin{equation}
{g_{hX_1^0X_1^0}} = \frac{1}{2}({a_{hX_1^0X_1^0}} + {b_{hX_1^0X_1^0}}) =  - \frac{2}{{\sqrt 3 }}({y_1}{\mathcal{N}_{21}} - {y_2}{\mathcal{N}_{31}}){\mathcal{N}_{11}}.
\end{equation}
In the zero momentum transfer limit, this induces a scalar interaction with
nucleons $N$:
\begin{equation}
{\mathcal{L}_{S,N}} = \sum\limits_{N = p,n} {{G_{S,N}}~\bar X_1^0X_1^0\bar NN},
\end{equation}
where
\begin{equation}
{G_{S,N}} =  - \frac{{{g_{hX_1^0X_1^0}}{m_N}}}{{2vm_h^2}}\left( {\sum\limits_{q = u,d,s} {f_q^N}  + 3f_Q^N} \right),
\end{equation}
and the nucleon form factors $f^N_i$ are determined to be roughly~\cite{Ellis:2000ds}:
\begin{eqnarray}
f_u^p &=& 0.020 \pm 0.004,~ f_d^p = 0.026 \pm 0.005,~ f_u^n = 0.014 \pm 0.003,
\nonumber\\
f_d^n &=& 0.036 \pm 0.008,~ f_s^p = f_s^n = 0.118 \pm 0.062,~ f_Q^N = \frac{2}{{27}}\left( {1 - \sum\limits_{q = u,d,s} {f_q^N} } \right).
\end{eqnarray}
The tiny up and down Yukawa couplings imply approximately iso-symmetric
couplings, ${G_{S,n}} \simeq {G_{S,p}}$, yielding
a spin-independent (SI) scattering cross section of
\begin{equation}
\sigma _{\chi N}^{{\mathrm{SI}}} = \frac{4}{\pi }\mu _{\chi N}^2G_{S,N}^2,
\label{eq:sigma_chiN:SI}
\end{equation}
where ${\mu _{\chi N}} \equiv {m_{\chi _1^0}}{m_N}/({m_{\chi _1^0}} + {m_N})$ is the $\chi _1^0$-$N$ reduced mass.

As a Majorana fermion, $\chi_1^0$ couples to $Z$ with an axial vector coupling of strength
\begin{equation}
{g_{ZX_1^0X_1^0}} = \frac{1}{2}({b_{ZX_1^0X_1^0}} - {a_{ZX_1^0X_1^0}}) = \frac{g}{{2{c_W}}}(|{\mathcal{N}_{31}}{|^2} - |{\mathcal{N}_{21}}|^2),
\end{equation}
leading to axial vector interactions with nucleons:
\begin{equation}
{\mathcal{L}_{A,N}} = \sum\limits_{N = p,n} {{G_{A,N}}~\bar X_1^0{\gamma ^\mu }{\gamma _5}X_1^0\bar N{\gamma ^\mu }{\gamma _5}N},
\end{equation}
where
\begin{equation}
{G_{A,q}} = \frac{{gg_A^q{g_{ZX_1^0X_1^0}}}}{{4{c_W}m_Z^2}}
~\text{and}~
{G_{A,N}} = \sum\limits_{q = u,d,s} {{G_{A,q}}} \Delta _q^N
~\text{with}~
g_A^u = \frac{1}{2}
~\text{and}~
g_A^d = g_A^s =  - \frac{1}{2}.
\end{equation}
The form factors are
$\Delta _u^p = \Delta _d^n = 0.842 \pm 0.012$, $\Delta _d^p = \Delta _u^n =  - 0.427 \pm 0.013$, 
$\Delta _s^p = \Delta _s^n =  - 0.085 \pm 0.018$~\cite{Airapetian:2006vy}.
These interactions lead to a spin-dependent (SD) scattering cross section
\begin{equation}
\sigma _{\chi N}^\mathrm{SD} = \frac{{12}}{\pi }\mu _{\chi N}^2G_{A,N}^2.
\label{eq:sigma_chiN:SD}
\end{equation}

Current limits on $\sigma_{\chi N}^\mathrm{SI}$ are lower than those on $\sigma_{\chi N}^\mathrm{SD}$ by several orders of magnitude
(owing to the coherent enhancement of the SI rate for heavy nuclear targets such as Xenon).
The green regions in Fig.~\ref{fig:limits:mQmT} are excluded by the 90\% CL exclusion limit on the 
SI DM-nucleon scattering cross section from LUX~\cite{Akerib:2013tjd}.
The profiles of these regions depend on the relation between $y_1$ and $y_2$.
As mentioned in Sec.~\ref{sec:custo}, in the custodial symmetry limit $g_{hX_1^0X_1^0}$ and $g_{ZX_1^0X_1^0}$ vanish for 
$m_Q<m_T$, while $g_{hX_1^0X_1^0}$ is nonzero for $m_T<m_Q$,
explaining why LUX only excludes the region with $m_T<m_Q$ in Fig.~\ref{fig:limits:mQmT:a} for $y_1 = y_2 = 0.5$.
In Figs.~\ref{fig:limits:mQmT:c} and \ref{fig:limits:mQmT:d}, the exclusion regions lying around the diagonals of the plots can reach
as high as $m_{\chi_1^0} \gtrsim 5~\TeV$,
because $g_{hX_1^0X_1^0}$ is enhanced when 
$m_Q \simeq m_T$ leads to comparable triplet and quadruplet components of $\chi_1^0$.

\subsection{Dark Matter Annihilation}

Finally, we consider bounds on the annihilation cross section $\langle \sigma_\mathrm{ann}v_\mathrm{rel} \rangle$ (where
$v_\mathrm{rel}$ is the relative velocity between two incoming DM particles) based on the non-observation of anomalous sources
of high energy gamma rays.
We adapt \texttt{MadGraph~5} to calculate the annihilation cross sections in the non-relativistic limit for all open two-body SM
final states.  The dominant channels\footnote{Annihilation into fermions mediated by the $Z$ or $h$ bosons
and into $hh$
are suppressed by either $v_\mathrm{rel}^2$, $m_f^2/m_{\chi_1^0}^2$, both \cite{Profumo:2013hqa}.} 
are $W^+W^-$, $ZZ$, and $Zh$.  The $W^+W^-$ channel
is typically dominant over $ZZ$ and $Zh$ by one to two orders of magnitude.

Thus, we compare predictions for annihilation into $W^+W^-$ with the null results for evidence of DM annihilation into gamma rays
in dwarf spheroidal galaxies based on 6 years of data collected by the {\em Fermi}-LAT
experiment~\cite{Ackermann:2015zua}.  {\em Fermi} provides 
$95\%$ CL upper limits on $\langle \sigma_\mathrm{ann}v_\mathrm{rel} \rangle$ for annihilation into $W^+W^-$ as a function of
the DM particle mass, which we translate into 
the exclusion regions shown as the red shaded regions in Fig.~\ref{fig:limits:mQmT}.
The {\em Fermi} data basically excludes 
$m_{\chi_1^0} \lesssim 1~\TeV$ for $m_Q<m_T$ and $m_{\chi_1^0} \lesssim 700~\GeV$ for $m_T<m_Q$.

\begin{figure}[!htbp]
\centering
\subfigure[$m_Q = 2.4~\TeV$, $m_T = 3~\TeV$.\label{fig:limits:y1y2:a}]
{\includegraphics[width=.49\textwidth]{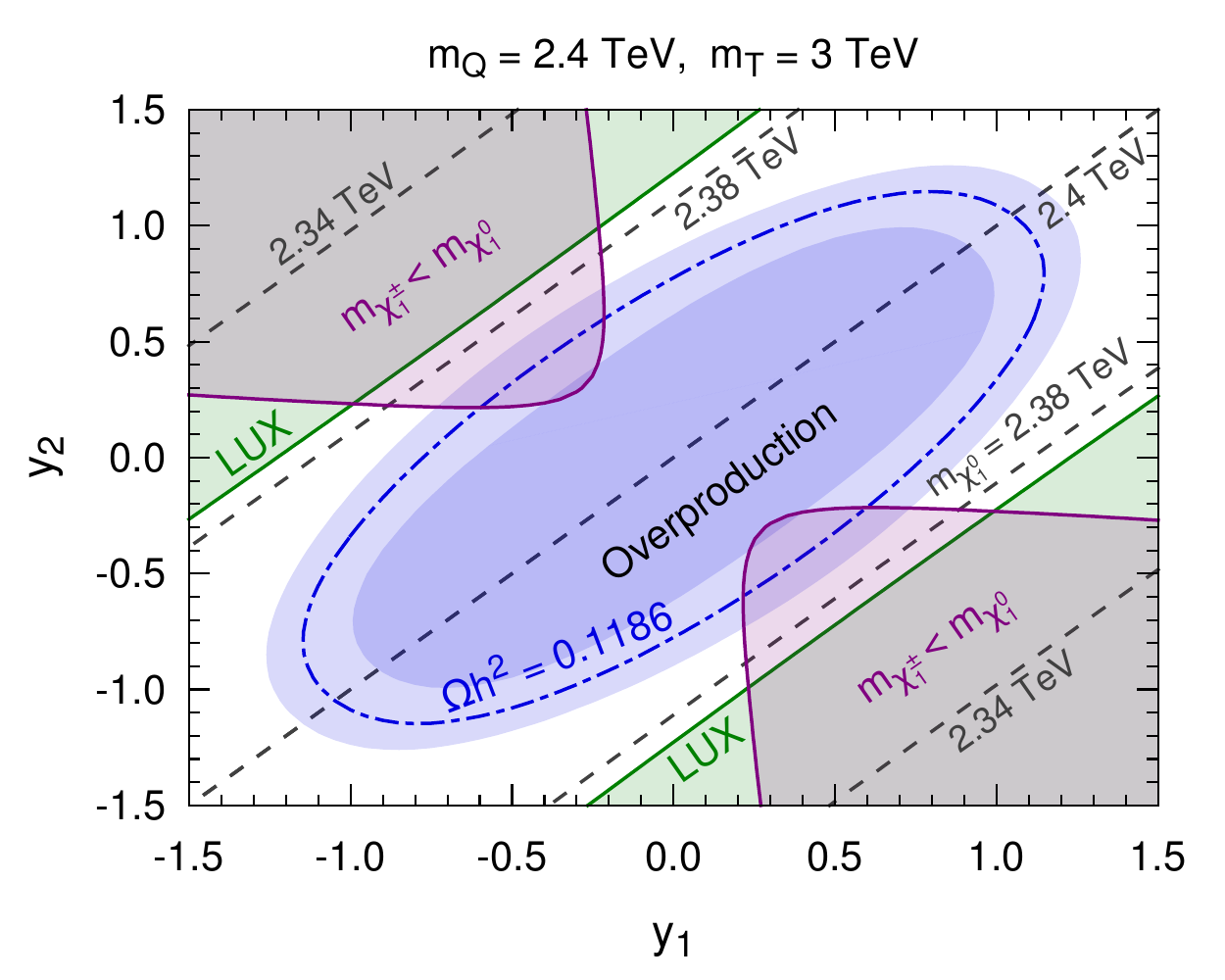}}
\subfigure[$m_Q = 3~\TeV$, $m_T = 2.4~\TeV$.\label{fig:limits:y1y2:b}]
{\includegraphics[width=.49\textwidth]{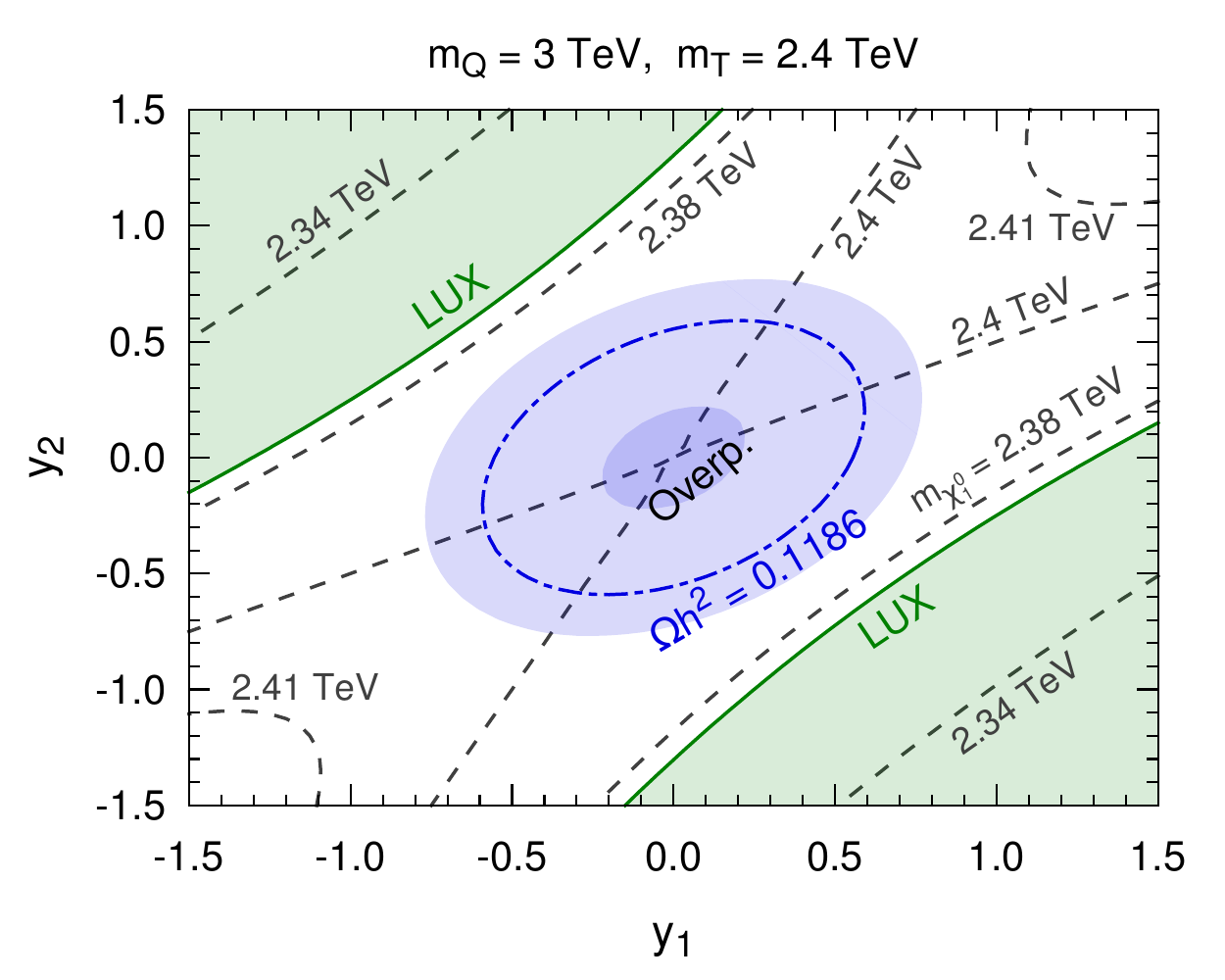}}
\subfigure[$m_Q = 1~\TeV$, $m_T = 1.7~\TeV$.\label{fig:limits:y1y2:c}]
{\includegraphics[width=.49\textwidth]{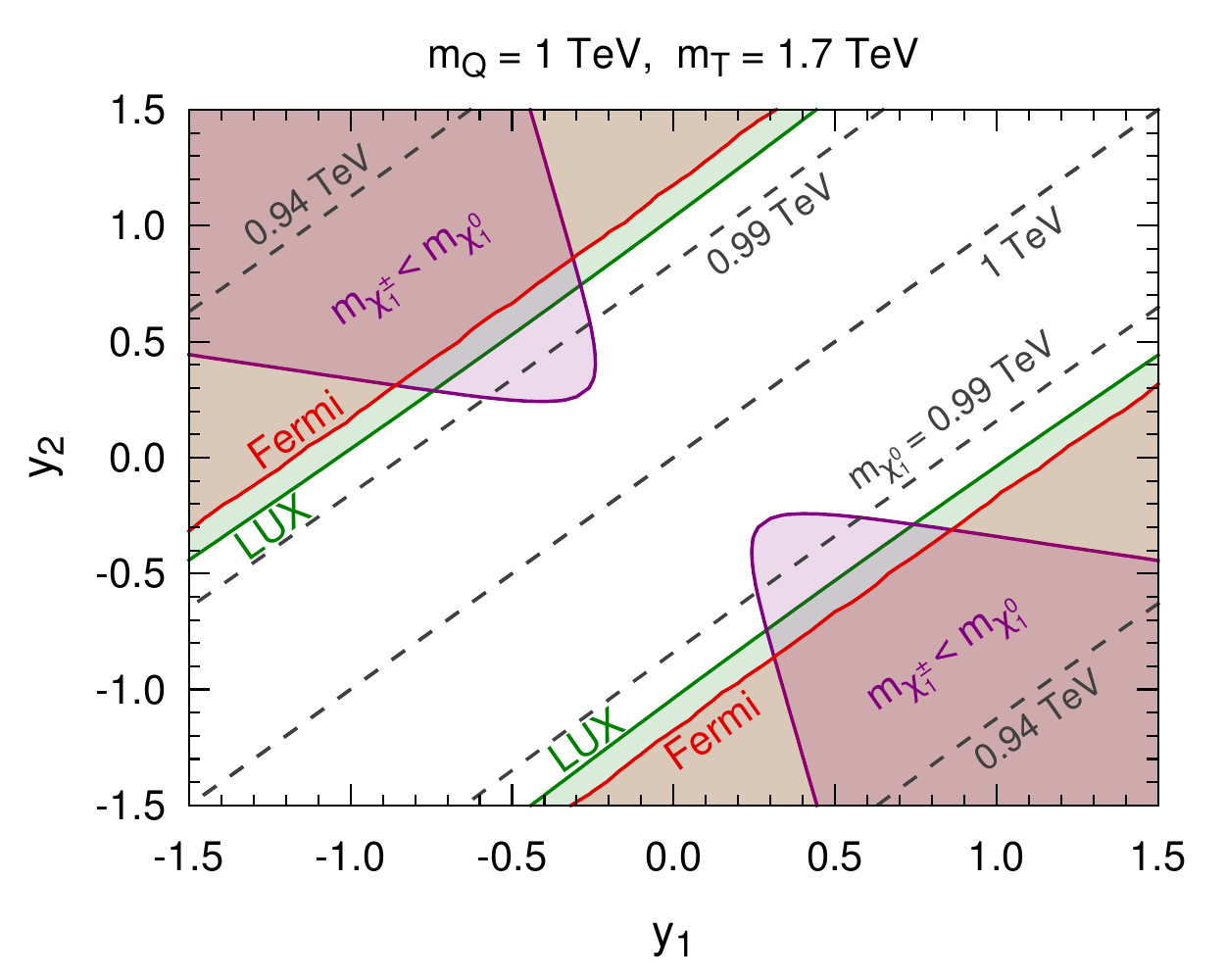}}
\subfigure[$m_Q = 1.5~\TeV$, $m_T = 0.7~\TeV$.\label{fig:limits:y1y2:d}]
{\includegraphics[width=.49\textwidth]{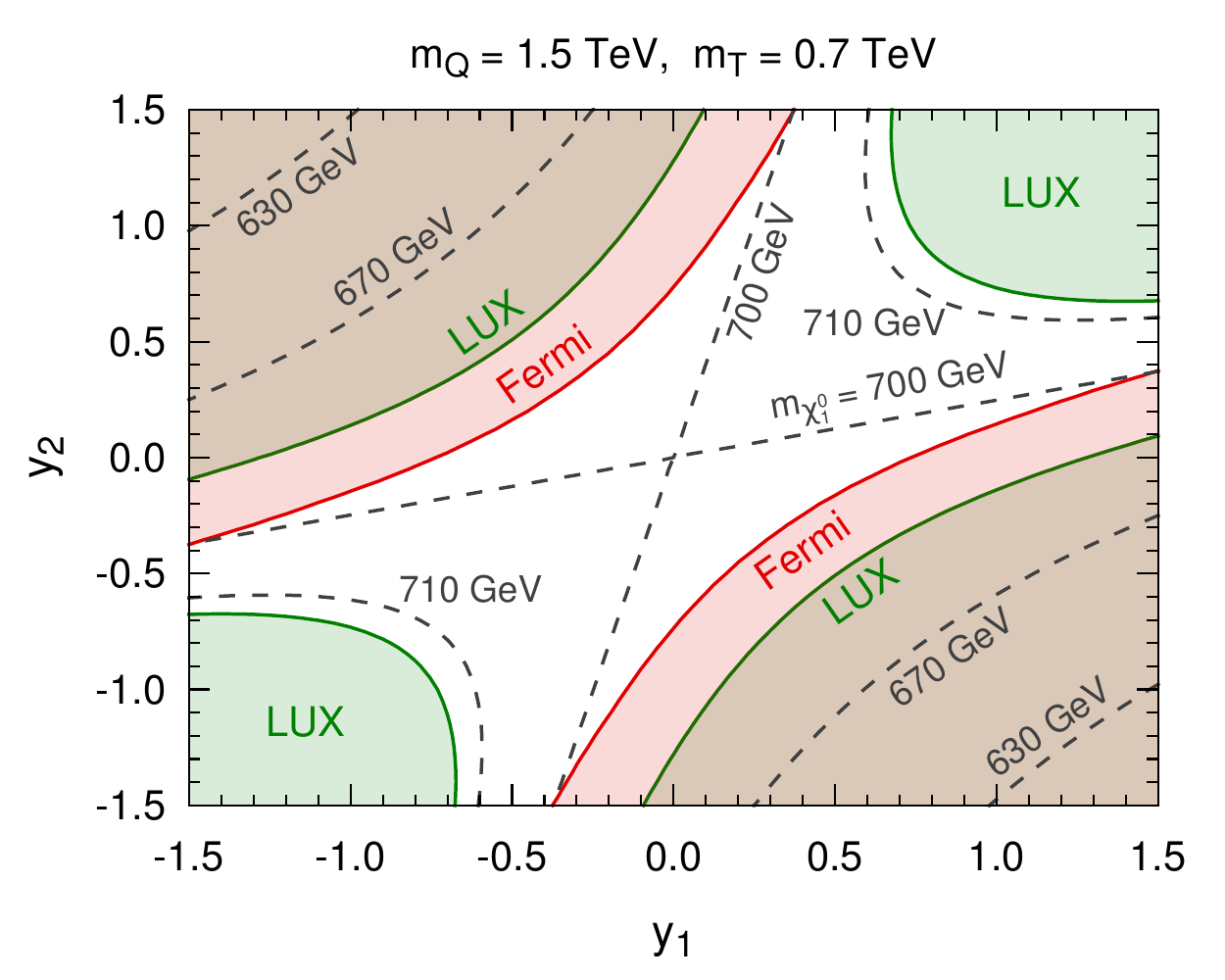}}
\caption{Constraints on the $y_1$-$y_2$ plane for four sets of fixed $m_Q$ and $m_T$ as indicated.
The legend for the lines and shadings are the same as in Fig.~\ref{fig:limits:mQmT}.}
\label{fig:limits:y1y2}
\end{figure}

\subsection{Constraints on the $y_1$-$y_2$ Plane}

By fixing the mass parameters $m_T$ and $m_Q$, we can see how the constraints vary in the $y_1$-$y_2$ plane, as 
shown in Fig.~\ref{fig:limits:y1y2}.
The plots are symmetric under the simultaneous transformations of $y_1 \to -y_1$ and $y_2 \to -y_2$.
In Figs.~\ref{fig:limits:y1y2:a} and \ref{fig:limits:y1y2:c}, we have $m_Q<m_T$, and the condition $m_{\chi_1^\pm} < m_{\chi_1^0}$ 
excludes some regions where $y_1$ and $y_2$ are sufficiently large and their signs are opposite to each other.
The contours of $m_{\chi_1^0}$ are parallel to the diagonals, which correspond to the custodial symmetry limit and 
have the largest values of $m_{\chi_1^0}$.
In Figs.~\ref{fig:limits:y1y2:b} and \ref{fig:limits:y1y2:d}, we have $m_T<m_Q$, and $m_{\chi_1^0}$ is larger at the corners of $y_1=y_2=1.5$ and $y_1=y_2=-1.5$ than at the corners of $y_1 = -y_2 = 1.5$ and $y_1 = -y_2 = -1.5$.

In Figs.~\ref{fig:limits:y1y2:a} and \ref{fig:limits:y1y2:b}, the fixed values of $m_T$ and $m_Q$ are 
suitable for obtaining an observed DM abundance.
The contours corresponding to the mean value of the measured $\Omega h^2$ appear as ellipses, inside which $\Omega h^2$ is larger.
The mass parameters are chosen to show  comparable sensitivities of LUX and \emph{Fermi}-LAT in Figs.~\ref{fig:limits:y1y2:c} 
and \ref{fig:limits:y1y2:d},
where both the LUX and \emph{Fermi} exclusion regions enclose the point $y_1 = -y_2 = 1.5$ as well as the point $y_1 = -y_2 = -1.5$.
In Fig.~\ref{fig:limits:y1y2:d}, the LUX bound also excludes the regions around $y_1=y_2=1.5$ and $y_1=y_2=-1.5$.
Both the LUX and \emph{Fermi} limits roughly coincide with the contours of $m_{\chi_1^0}$.

\section{Conclusions and Outlook}
\label{sec:concl}

In this work, we explore a dark sector consisting of a fermionic $SU(2)_L$ triplet and two fermionic $SU(2)_L$ quadruplets.
This set-up is a minimal UV complete realistic model of electroweakly interacting dark matter with tree level coupling to the SM
Higgs boson, the simplest such construction which is distinct from any limit of the MSSM.
After electroweak symmetry-breaking, the dark sector consists of
three Majorana fermions $\chi_i^0$, three singly charged fermions $\chi_i^\pm$, and one doubly charged fermion $\chi^{\pm\pm}$.
The lightest neutral fermion $\chi_1^0$ is a wonderful DM candidate provided it is the lightest of the dark sector fermions.

When two Yukawa couplings are equal, i.e., $y_1=y_2$, there is an approximate global custodial symmetry, implying that
$\chi_i^0$ is mass-degenerate with $\chi_i^\pm$ at tree level.
We compute the one-loop mass corrections to determine the precise spectrum.
Fortunately, in the custodial limit these corrections always increase the masses of charged fermions.
Another gift from this symmetry is the tree-level vanishing of the $\chi_1^0$ couplings to $Z$ and $h$,
rendering the current DM direct searches impotent as far as constraining it.
Beyond the custodial symmetry limit, at tree level $m_{\chi_i^0}$ and $m_{\chi_i^\pm}$ are slightly different, but 
nonetheless still quite degenerate.
At the one-loop level, mass corrections suggest that we may have $m_{\chi_1^\pm}<m_{\chi_1^0}$ when 
$y_1$ and $y_2$ have opposite signs.
When that happens, $\chi_1^0$ is no longer a viable DM candidate, decaying into the lightest charged state.

Due to the mass degeneracy, coannihilation processes among dark sector fermions strongly affect the 
abundance evolution of $\chi_1^0$ in the early Universe, and must be included.
The calculation suggests that $m_{\chi_1^0} \sim 2.4~\TeV$ to saturate the observed relic density for
a standard cosmology.
We also investigate the constraints from the electroweak oblique parameters and direct and indirect searches.
The global fit result of $S$, $T$, and $U$ parameters excludes a region up to 
$m_Q\lesssim 300~(400)~\GeV$ and $m_T\lesssim 1.8~(4.1)~\TeV$ for $y_1=y_2=0.5$ ($y_1=0.5$ and $y_2=1$).
The LUX exclusion region significantly depends on the relation between $y_1$ and $y_2$.
When $y_2$ has a sign opposite to $y_1$, the LUX result excludes $m_{\chi_1^0}$ up to several TeV for $m_Q \simeq m_T$, 
cutting in to some regions favored by the relic abundance.
Annihilation into $W^+W^-$ is the dominant channel in the non-relativistic limit, and
{\em Fermi}-LAT dwarf galaxy limits
exclude $m_{\chi_1^0}$ $\lesssim 1~\TeV$ and $\lesssim 700~\GeV$ for $m_Q < m_T$ and $m_T < m_Q$, respectively.
Nonetheless, there is still plenty of room in the parameter space that is consistent with the observed DM abundance and escaping from phenomenological constraints.

As the charged fermions in the dark sector couple to the Higgs boson, the $h\to\gamma\gamma$ decay is a possible indirect probe of
their presence.
However, the current LHC data are not sufficiently precise to give a meaningful limit, though LHC high luminosity
running may reach the correct ballpark \cite{Dawson:2013bba}.
LHC direct searches for exotic charged particles decaying into missing momentum may also be able to explore the model, 
but the electroweak production rates of the dark sector charged fermions are quite low for multi-TeV fermions, and it may
ultimately fall to future higher energy colliders to have the last word \cite{Low:2014cba,Bramante:2014tba}.

\acknowledgments

TMPT acknowledges Randy Cotta, JoAnne Hewett, and Devin Walker for earlier collaboration on related topics.
ZHY would like to acknowledge helpful discussions with Arvind Rajaraman, Philip Tanedo, Alexander Wijangco, 
Mohammad Abdullah, and Ye-Ling Zhou and also thanks the Particle Theory Group at UC Irvine 
for hospitality during his visit.
The work of TMPT is supported in part by NSF grant PHY-1316792 and by the University of California, Irvine through a Chancellor's Fellowship,
and that of ZHY was supported by the China Scholarship Council (Grant No. 201404910374).

\appendix

\section{Detailed Expressions for Interaction Terms}
\label{app:int}

In this appendix, we derive explicit expressions for the interaction terms in the triplet-quadruplet model.
The covariant derivatives for the triplet and quadruplets are
\begin{eqnarray}
{D_\mu }T &=& ({\partial _\mu } - igW_\mu ^at_T^a)T,
\\
{D_\mu }{Q_i} &=& ({\partial _\mu } - ig'{B_\mu }{Y_{{Q_i}}} - igW_\mu ^at_Q^a){Q_i},
\end{eqnarray}
where $Y_{Q_1} = -1/2$, $Y_{Q_2} = +1/2$, and the generators of $SU(2)_L$ are:
\begin{equation}
\setlength\arraycolsep{0.5em}
t_T^1 = \frac{1}{{\sqrt 2 }}\left( {\begin{array}{*{20}{c}}
   {} & 1 & {}  \\
   1 & {} & { - 1}  \\
   {} & { - 1} & {}  \\
 \end{array} } \right),~
t_T^2 = \frac{1}{{\sqrt 2 }}\left( {\begin{array}{*{20}{c}}
   {} & { - i} & {}  \\
   i & {} & i  \\
   {} & { - i} & {}  \\
 \end{array} } \right),~
t_T^3 = \left( {\begin{array}{*{20}{c}}
   1 & {} & {}  \\
   {} & 0 & {}  \\
   {} & {} & { - 1}  \\
 \end{array} } \right),
\end{equation}
and
\begin{eqnarray}
&& t_Q^1 = \left( {\begin{array}{*{20}{c}}
   {} & {\sqrt 3 /2} & {} & {}  \\
   {\sqrt 3 /2} & {} & 1 & {}  \\
   {} & 1 & {} & {\sqrt 3 /2}  \\
   {} & {} & {\sqrt 3 /2} & {}  \\
 \end{array} } \right),~
t_Q^2 = \left( {\begin{array}{*{20}{c}}
   {} & { - \sqrt 3 i/2} & {} & {}  \\
   {\sqrt 3 i/2} & {} & { - i} & {}  \\
   {} & i & {} & { - \sqrt 3 i/2}  \\
   {} & {} & {\sqrt 3 i/2} & {}  \\
 \end{array} } \right),
\nonumber\\
&&\hspace*{11em}
t_Q^3 = {\mathrm{diag}}\left( {\frac{3}{2},\frac{1}{2}, - \frac{1}{2}, - \frac{3}{2}} \right).
\end{eqnarray}
We can express the gauge interaction terms in Eqs.~\eqref{eq:Lag_T} and \eqref{eq:Lag_Q} as
\begin{eqnarray}
{\mathcal{L}_{\mathrm{T}}} &\supset& {T^\dag }{{\bar \sigma }^\mu }gW_\mu ^at_T^aT
\nonumber\\
 &=& (e{A_\mu } + g{c_W}{Z_\mu }){({T^ + })^\dag }{{\bar \sigma }^\mu }{T^ + } + gW_\mu ^ + {({T^ + })^\dag }{{\bar \sigma }^\mu }{T^0}
 + gW_\mu ^ - {({T^0})^\dag }{{\bar \sigma }^\mu }{T^ + }
\nonumber\\
&& - gW_\mu ^ + {({T^0})^\dag }{{\bar \sigma }^\mu }{T^ - }
 - gW_\mu ^ - {({T^ - })^\dag }{{\bar \sigma }^\mu }{T^0} - (e{A_\mu } + g{c_W}{Z_\mu }){({T^ - })^\dag }{{\bar \sigma }^\mu }{T^ - }
\label{eq:Lag_T:int}
\end{eqnarray}
and
\begin{eqnarray}
{\mathcal{L}_{\mathrm{Q}}} &\supset& Q_1^\dag {{\bar \sigma }^\mu }(g'{B_\mu }{Y_{{Q_1}}} + gW_\mu ^at_Q^a){Q_1} + Q_2^\dag {{\bar \sigma }^\mu }(g'{B_\mu }{Y_{{Q_2}}} + gW_\mu ^at_Q^a){Q_2}
\nonumber\\
 &=& \frac{{\sqrt 6 }}{2}gW_\mu ^ + [{(Q_1^ + )^\dag }{{\bar \sigma }^\mu }Q_1^0 + {(Q_2^{ +  + })^\dag }{{\bar \sigma }^\mu }Q_2^ + ] + \frac{{\sqrt 6 }}{2}gW_\mu ^ - [{(Q_1^0)^\dag }{{\bar \sigma }^\mu }Q_1^ +  + {(Q_2^ + )^\dag }{{\bar \sigma }^\mu }Q_2^{ +  + }]
\nonumber\\
&& + \sqrt 2 gW_\mu ^ + [{(Q_1^0)^\dag }{{\bar \sigma }^\mu }Q_1^ -  + {(Q_2^ + )^\dag }{{\bar \sigma }^\mu }Q_2^0] + \sqrt 2 gW_\mu ^ - [{(Q_1^ - )^\dag }{{\bar \sigma }^\mu }Q_1^0 + {(Q_2^0)^\dag }{{\bar \sigma }^\mu }Q_2^ + ]
\nonumber\\
&& + \frac{{\sqrt 6 }}{2}gW_\mu ^ + [{(Q_1^ - )^\dag }{{\bar \sigma }^\mu }Q_1^{ -  - } + {(Q_2^0)^\dag }{{\bar \sigma }^\mu }Q_2^ - ] + \frac{{\sqrt 6 }}{2}gW_\mu ^ - [{(Q_1^{ -  - })^\dag }{{\bar \sigma }^\mu }Q_1^ -  + {(Q_2^ - )^\dag }{{\bar \sigma }^\mu }Q_2^0]
\nonumber\\
&& + \frac{g}{{2{c_W}}}{Z_\mu }{(Q_1^0)^\dag }{{\bar \sigma }^\mu }Q_1^0 - \frac{g}{{2{c_W}}}{Z_\mu }{(Q_2^0)^\dag }{{\bar \sigma }^\mu }Q_2^0
\nonumber\\
&& + \left[ {e{A_\mu } + \frac{{g(s_W^2 + 3c_W^2)}}{{2{c_W}}}{Z_\mu }} \right]{(Q_1^ + )^\dag }{{\bar \sigma }^\mu }Q_1^ +  + \left[ {e{A_\mu } + \frac{{g(c_W^2 - s_W^2)}}{{2{c_W}}}{Z_\mu }} \right]{(Q_2^ + )^\dag }{{\bar \sigma }^\mu }Q_2^ + 
\nonumber\\
&& + \left[ { - e{A_\mu } + \frac{{g(s_W^2 - c_W^2)}}{{2{c_W}}}{Z_\mu }} \right]{(Q_1^ - )^\dag }{{\bar \sigma }^\mu }Q_1^ -  + \left[ { - e{A_\mu } - \frac{{g(3c_W^2 + s_W^2)}}{{2{c_W}}}{Z_\mu }} \right]{(Q_2^ - )^\dag }{{\bar \sigma }^\mu }Q_2^ - 
\nonumber\\
&& + \left[ { - 2e{A_\mu } + \frac{{g(s_W^2 - 3c_W^2)}}{{2{c_W}}}{Z_\mu }} \right]{(Q_1^{ -  - })^\dag }{{\bar \sigma }^\mu }Q_1^{ -  - }
\nonumber\\
&& + \left[ {2e{A_\mu } + \frac{{g(3c_W^2 - s_W^2)}}{{2{c_W}}}{Z_\mu }} \right]{(Q_2^{ +  + })^\dag }{{\bar \sigma }^\mu }Q_2^{ +  + }.
\label{eq:Lag_Q:int}
\end{eqnarray}
Including the would-be Goldstone bosons, Eq.~\eqref{eq:Lag_HTQ} becomes
\begin{eqnarray}
{\mathcal{L}_{{\mathrm{HTQ}}}} &=& {y_1}{G^ + }\left( {Q_1^{ -  - }{T^ + } - \frac{2}{{\sqrt 6 }}Q_1^ - {T^0} - \frac{1}{{\sqrt 3 }}Q_1^0{T^ - }} \right)
\nonumber\\
&& + {y_1}(v + h + i{G^0})\left( {\frac{1}{{\sqrt 6 }}Q_1^ - {T^ + } - \frac{1}{{\sqrt 3 }}Q_1^0{T^0} - \frac{1}{{\sqrt 2 }}Q_1^ + {T^ - }} \right)
\nonumber\\
&& + {y_2}{G^ - }\left( { - Q_2^{ +  + }{T^ - } - \frac{2}{{\sqrt 6 }}Q_2^ + {T^0} + \frac{1}{{\sqrt 3 }}Q_2^0{T^ + }} \right)
\nonumber\\
&& + {y_2}(v + h - i{G^0})\left( {\frac{1}{{\sqrt 3 }}Q_2^0{T^0} + \frac{1}{{\sqrt 6 }}Q_2^ + {T^ - } - \frac{1}{{\sqrt 2 }}Q_2^ - {T^ + }} \right) + \mathrm{h.c.}\,,
\label{eq:Lag_HTQ:int}
\end{eqnarray}
where the Goldstone bosons $G^0$ and $G^\pm$ are defined as
\begin{equation}
H = \left( {\begin{array}{*{20}{c}}
   {{H^ + }}  \\
   {{H^0}}  \\
 \end{array} } \right) = \left( {\begin{array}{*{20}{c}}
   {{G^ + }}  \\
   {\dfrac{1}{{\sqrt 2 }}(v + h + i{G^0})}  \\
 \end{array} } \right).
\end{equation}

For convenience, we would like to express the interaction terms with 4-component fermionic fields.
Here we define
\begin{equation}
\Psi _i^0 = \left( {\begin{array}{*{20}{c}}
   {\psi _{iL}^0}  \\
   {{{(\psi _{iR}^0)}^\dag }}  \\
 \end{array} } \right),~
\Psi _i^ +  = \left( {\begin{array}{*{20}{c}}
   {\psi _{iL}^ + }  \\
   {{{(\psi _{iR}^ - )}^\dag }}  \\
 \end{array} } \right),
\end{equation}
where
\begin{equation}
\psi _L^0 = \psi _R^0 = {({T^0},Q_1^0,Q_2^0)^{\mathrm{T}}},~
\psi _L^ +  = {({T^ + },Q_1^ + ,Q_2^ + )^{\mathrm{T}}},~
\psi _R^ -  = {({T^ - },Q_1^ - ,Q_2^ - )^{\mathrm{T}}}.
\end{equation}
Now Eq.~\eqref{eq:state_mix} is equivalent to
\begin{equation}
\psi _{L,R}^0 = \mathcal{N}\chi _{L,R}^0,~
\psi _L^ +  = {\mathcal{C}_L}\chi _L^ + ,~
\psi _R^ -  = {\mathcal{C}_R}\chi _R^ - .
\end{equation}
We can use chiral projection operators to divide every fermionic field into two parts:
\begin{equation}
\Psi _{iL}^{0, + } = {P_L}\Psi _i^{0, + },~
\Psi _{iR}^{0, + } = {P_R}\Psi _i^{0, + },~
X_{iL}^{0, + , +  + } = {P_L}X_i^{0, + , +  + },~
X_{iR}^{0, + , +  + } = {P_R}X_i^{0, + , +  + }.
\end{equation}
Thus we have
\begin{equation}
\Psi _{iL}^0 = {\mathcal{N}_{ij}}X_{jL}^0,~
\Psi _{iR}^0 = \mathcal{N}_{ij}^*X_{jR}^0,~
\Psi _{iL}^ +  = {({\mathcal{C}_L})_{ij}}X_{jL}^ + ,~
\Psi _{iR}^ +  = ({\mathcal{C}_R})_{ij}^*X_{jR}^ + .
\end{equation}

The interaction terms in Eqs.~\eqref{eq:Lag_T:int}, \eqref{eq:Lag_Q:int}, and \eqref{eq:Lag_HTQ:int} can be written down with the 4-component fields $\Psi_{i}^0$ and $\Psi_{i}^+$ projected into their left- and right-handed parts:
\begin{eqnarray}
{\mathcal{L}_\Psi } &=& {a_{A\Psi _i^ + \Psi _i^ + }}{A_\mu }\bar \Psi _{iL}^ + {\gamma ^\mu }\Psi _{iL}^ +  + {b_{A\Psi _i^ + \Psi _i^ + }}{A_\mu }\bar \Psi _{iR}^ + {\gamma ^\mu }\Psi _{iR}^ +  + {a_{Z\Psi _i^ + \Psi _i^ + }}{Z_\mu }\bar \Psi _{iL}^ + {\gamma ^\mu }\Psi _{iL}^ + 
\nonumber\\
&& + {b_{Z\Psi _i^ + \Psi _i^ + }}{Z_\mu }\bar \Psi _{iR}^ + {\gamma ^\mu }\Psi _{iR}^ +  + \frac{1}{2}({a_{Z\Psi _i^0\Psi _i^0}}{Z_\mu }\bar \Psi _{iL}^0{\gamma ^\mu }\Psi _{iL}^0 + {b_{Z\Psi _i^0\Psi _i^0}}{Z_\mu }\bar \Psi _{iR}^0{\gamma ^\mu }\Psi _{iR}^0)
\nonumber\\
&& + {a_{W\Psi _i^ + \Psi _i^0}}(W_\mu ^ + \bar \Psi _{iL}^ + {\gamma ^\mu }\Psi _{iL}^0 + \mathrm{h.c.}) + {b_{W\Psi _i^ + \Psi _i^0}}(W_\mu ^ + \bar \Psi _{iR}^ + {\gamma ^\mu }\Psi _{iR}^0 + \mathrm{h.c.})
\nonumber\\
&& + {a_{W{X^{ +  + }}\Psi _i^ + }}(W_\mu ^ + \bar X_L^{ +  + }{\gamma ^\mu }\Psi _{iL}^ +  + \mathrm{h.c.}) + {b_{W{X^{ +  + }}\Psi _i^ + }}(W_\mu ^ + \bar X_R^{ +  + }{\gamma ^\mu }\Psi _{iR}^ +  + \mathrm{h.c.})
\nonumber\\
&& + \frac{1}{2}({a_{h\Psi _i^0\Psi _j^0}}h\bar \Psi _{iR}^0\Psi _{jL}^0 + {b_{h\Psi _i^0\Psi _j^0}}h\bar \Psi _{iL}^0\Psi _{jR}^0 + {a_{{G^0}\Psi _i^0\Psi _j^0}}{G^0}\bar \Psi _{iR}^0\Psi _{jL}^0 + {b_{{G^0}\Psi _i^0\Psi _j^0}}{G^0}\bar \Psi _{iL}^0\Psi _{jR}^0)
\nonumber\\
&& + {a_{h\Psi _i^ + \Psi _j^ + }}h\bar \Psi _{iR}^ + \Psi _{jL}^ +  + {b_{h\Psi _i^ + \Psi _j^ + }}h\bar \Psi _{iL}^ + \Psi _{jR}^ +  + {a_{{G^0}\Psi _i^ + \Psi _j^ + }}{G^0}\bar \Psi _{iR}^ + \Psi _{jL}^ +  + {b_{{G^0}\Psi _i^ + \Psi _j^ + }}{G^0}\bar \Psi _{iL}^ + \Psi _{jR}^ + 
\nonumber\\
&& + {a_{{G^ \pm }\Psi _i^ + \Psi _j^0}}({G^ + }\bar \Psi _{iR}^ + \Psi _{jL}^0 + \mathrm{h.c.}) + {b_{{G^ \pm }\Psi _i^ + \Psi _j^0}}({G^ + }\bar \Psi _{iL}^ + \Psi _{jR}^0 + \mathrm{h.c.})
\nonumber\\
&& + {a_{{G^ \pm }{X^{ +  + }}\Psi _i^ + }}({G^ + }\bar X_R^{ +  + }\Psi _{iL}^ +  + \mathrm{h.c.}) + {b_{{G^ \pm }{X^{ +  + }}\Psi _i^ + }}({G^ + }\bar X_L^{ +  + }\Psi _{iR}^ +  + \mathrm{h.c.}),
\end{eqnarray}
where the coupling coefficients read
\begin{eqnarray}
&& {a_{A\Psi _1^ + \Psi _1^ + }} = {b_{A\Psi _1^ + \Psi _1^ + }} = {a_{A\Psi _2^ + \Psi _2^ + }} = {b_{A\Psi _2^ + \Psi _2^ + }} = {a_{A\Psi _3^ + \Psi _3^ + }} = {b_{A\Psi _3^ + \Psi _3^ + }} = e,
\nonumber\\
&& {a_{Z\Psi _1^ + \Psi _1^ + }} = {b_{Z\Psi _1^ + \Psi _1^ + }} = g{c_W},~ {a_{Z\Psi _2^ + \Psi _2^ + }} = \frac{g}{{2{c_W}}}(3c_W^2 + s_W^2) = {b_{Z\Psi _3^ + \Psi _3^ + }},
\nonumber\\
&& {b_{Z\Psi _2^ + \Psi _2^ + }} = \frac{g}{{2{c_W}}}(c_W^2 - s_W^2) = {a_{Z\Psi _3^ + \Psi _3^ + }},
\nonumber\\
&& {a_{Z\Psi _2^0\Psi _2^0}} =  - {b_{Z\Psi _2^0\Psi _2^0}} = \frac{g}{{2{c_W}}},~ {a_{Z\Psi _3^0\Psi _3^0}} =  - {b_{Z\Psi _3^0\Psi _3^0}} =  - \frac{g}{{2{c_W}}},
\nonumber\\
&& {a_{W\Psi _1^ + \Psi _1^0}} = {b_{W\Psi _1^ + \Psi _1^0}} = g,~ {a_{W\Psi _2^ + \Psi _2^0}} = \frac{{\sqrt 6 }}{2}g =  - {b_{W\Psi _3^ + \Psi _3^0}},~ {b_{W\Psi _2^ + \Psi _2^0}} =  - \sqrt 2 g =  - {a_{W\Psi _3^ + \Psi _3^0}},
\nonumber\\
&& {b_{W{X^{ +  + }}\Psi _2^ + }} =  - \frac{{\sqrt 6 }}{2}g,~
{a_{W{X^{ +  + }}\Psi _3^ + }} = \frac{{\sqrt 6 }}{2}g,
\nonumber\\
&& {a_{h\Psi _1^0\Psi _2^0}} = {b_{h\Psi _1^0\Psi _2^0}} =  - \frac{{{y_1}}}{{\sqrt 3 }} = {a_{h\Psi _2^0\Psi _1^0}} = {b_{h\Psi _2^0\Psi _1^0}},
\nonumber\\
&& {a_{h\Psi _1^0\Psi _3^0}} = {b_{h\Psi _1^0\Psi _3^0}} = \frac{{{y_2}}}{{\sqrt 3 }} = {a_{h\Psi _3^0\Psi _1^0}} = {b_{h\Psi _3^0\Psi _1^0}},
\nonumber\\
&& {a_{{G^0}\Psi _1^0\Psi _2^0}} =  - {b_{{G^0}\Psi _1^0\Psi _2^0}} =  - \frac{{{y_1}}}{{\sqrt 3 }}i = {a_{{G^0}\Psi _2^0\Psi _1^0}} =  - {b_{{G^0}\Psi _2^0\Psi _1^0}},
\nonumber\\
&&  {a_{{G^0}\Psi _1^0\Psi _3^0}} =  - {b_{{G^0}\Psi _1^0\Psi _3^0}} =  - \frac{{{y_2}}}{{\sqrt 3 }}i = {a_{{G^0}\Psi _3^0\Psi _1^0}} =  - {b_{{G^0}\Psi _3^0\Psi _1^0}},
\nonumber\\
&& {a_{h\Psi _1^ + \Psi _2^ + }} = {b_{h\Psi _2^ + \Psi _1^ + }} =  - \frac{{{y_1}}}{{\sqrt 2 }},~ {a_{h\Psi _2^ + \Psi _1^ + }} = {b_{h\Psi _1^ + \Psi _2^ + }} = \frac{{{y_1}}}{{\sqrt 6 }},
\nonumber\\
&& {a_{h\Psi _1^ + \Psi _3^ + }} = {b_{h\Psi _3^ + \Psi _1^ + }} = \frac{{{y_2}}}{{\sqrt 6 }},~ {a_{h\Psi _3^ + \Psi _1^ + }} = {b_{h\Psi _1^ + \Psi _3^ + }} =  - \frac{{{y_2}}}{{\sqrt 2 }},
\nonumber\\
&& {a_{{G^0}\Psi _1^ + \Psi _2^ + }} =  - {b_{{G^0}\Psi _2^ + \Psi _1^ + }} =  - \frac{{{y_1}}}{{\sqrt 2 }}i,~ {a_{{G^0}\Psi _2^ + \Psi _1^ + }} =  - {b_{{G^0}\Psi _1^ + \Psi _2^ + }} = \frac{{{y_1}}}{{\sqrt 6 }}i,
\nonumber\\
&& {a_{{G^0}\Psi _1^ + \Psi _3^ + }} =  - {b_{{G^0}\Psi _3^ + \Psi _1^ + }} =  - \frac{{{y_2}}}{{\sqrt 6 }}i,~ {a_{{G^0}\Psi _3^ + \Psi _1^ + }} =  - {b_{{G^0}\Psi _1^ + \Psi _3^ + }} = \frac{{{y_2}}}{{\sqrt 2 }}i,
\nonumber\\
&& {a_{{G^ \pm }\Psi _1^ + \Psi _2^0}} =  - \frac{{{y_1}}}{{\sqrt 3 }},~ {a_{{G^ \pm }\Psi _2^ + \Psi _1^0}} =  - \frac{{2{y_1}}}{{\sqrt 6 }},~ {b_{{G^ \pm }\Psi _1^ + \Psi _3^0}} = \frac{{{y_2}}}{{\sqrt 3 }},~ {b_{{G^ \pm }\Psi _3^ + \Psi _1^0}} =  - \frac{{2{y_2}}}{{\sqrt 6 }},
\nonumber\\
&& {a_{{G^ \pm }{X^{ +  + }}\Psi _1^ + }} = {y_1},~ {b_{{G^ \pm }{X^{ +  + }}\Psi _1^ + }} =  - {y_2}.
\label{eq:coeff_gauge}
\end{eqnarray}
The coupling coefficients that have not mentioned above are zero.
Converting the gauge bases into the physical bases, we have
\begin{eqnarray}
{\mathcal{L}_X}&&\hspace*{-0.4em} = {a_{AX_i^ + X_j^ + }}{A_\mu }\bar X_{iL}^ + {\gamma ^\mu }X_{jL}^ +  + {b_{AX_i^ + X_j^ + }}{A_\mu }\bar X_{iR}^ + {\gamma ^\mu }X_{jR}^ +  + {a_{ZX_i^ + X_j^ + }}{Z_\mu }\bar X_{iL}^ + {\gamma ^\mu }X_{jL}^ + 
\nonumber\\
&& + {b_{ZX_i^ + X_j^ + }}{Z_\mu }\bar X_{iR}^ + {\gamma ^\mu }X_{jR}^ +  + \frac{1}{2}({a_{ZX_i^0X_j^0}}{Z_\mu }\bar X_{iL}^0{\gamma ^\mu }X_{jL}^0 + {b_{ZX_i^0X_j^0}}{Z_\mu }\bar X_{iR}^0{\gamma ^\mu }X_{jR}^0)
\nonumber\\
&& + {a_{A{X^{ +  + }}{X^{ +  + }}}}{A_\mu }\bar X_L^{ +  + }{\gamma ^\mu }X_L^{ +  + } + {b_{A{X^{ +  + }}{X^{ +  + }}}}{A_\mu }\bar X_R^{ +  + }{\gamma ^\mu }X_R^{ +  + }
\nonumber\\
&& + {a_{Z{X^{ +  + }}{X^{ +  + }}}}{Z_\mu }\bar X_L^{ +  + }{\gamma ^\mu }X_L^{ +  + } + {b_{Z{X^{ +  + }}{X^{ +  + }}}}{Z_\mu }\bar X_R^{ +  + }{\gamma ^\mu }X_R^{ +  + }
\nonumber\\
&& + ({a_{WX_i^ + X_j^0}}W_\mu ^ + \bar X_{iL}^ + {\gamma ^\mu }X_{jL}^0 + \mathrm{h.c.}) + ({b_{WX_i^ + X_j^0}}W_\mu ^ + \bar X_{iR}^ + {\gamma ^\mu }X_{jR}^0 + \mathrm{h.c.})
\nonumber\\
&& + ({a_{W{X^{ +  + }}X_i^ + }}W_\mu ^ + \bar X_L^{ +  + }{\gamma ^\mu }X_{iL}^ +  + \mathrm{h.c.}) + ({b_{W{X^{ +  + }}X_i^ + }}\bar X_R^{ +  + }{\gamma ^\mu }X_{iR}^ +  + \mathrm{h.c.})
\nonumber\\
&& + \frac{1}{2}({a_{hX_i^0X_j^0}}h\bar X_{iR}^0X_{jL}^0 + {b_{hX_i^0X_j^0}}h\bar X_{iL}^0X_{jR}^0 + {a_{{G^0}X_i^0X_j^0}}{G^0}\bar X_{iR}^0X_{jL}^0 + {b_{{G^0}X_i^0X_j^0}}{G^0}\bar X_{iL}^0X_{jR}^0)
\nonumber\\
&& + {a_{hX_i^ + X_j^ + }}h\bar X_{iR}^ + X_{jL}^ +  + {b_{hX_i^ + X_j^ + }}h\bar X_{iL}^ + X_{jR}^ +  + {a_{{G^0}X_i^ + X_j^ + }}{G^0}\bar X_{iR}^ + X_{jL}^ +  + {b_{{G^0}X_i^ + X_j^ + }}{G^0}\bar X_{iL}^ + X_{jR}^ + 
\nonumber\\
&& + ({a_{{G^ \pm }X_i^ + X_j^0}}{G^ + }\bar X_{iR}^ + X_{jL}^0 + \mathrm{h.c.}) + ({b_{{G^ \pm }X_i^ + X_j^0}}{G^ + }\bar X_{iL}^ + X_{jR}^0 + \mathrm{h.c.})
\nonumber\\
&& + ({a_{{G^ \pm }{X^{ +  + }}X_i^ + }}{G^ + }\bar X_R^{ +  + }X_{iL}^ +  + \mathrm{h.c.}) + ({b_{{G^ \pm }{X^{ +  + }}X_i^ + }}{G^ + }\bar X_L^{ +  + }X_{iR}^ +  + \mathrm{h.c.}),
\end{eqnarray}
where the coupling coefficients are related to those in \eqref{eq:coeff_gauge} through the mixing matrices:
\begin{eqnarray}
&& {a_{AX_i^ + X_j^ + }} = {a_{A\Psi _k^ + \Psi _k^ + }}({\mathcal{C}_L})_{ki}^*{({\mathcal{C}_L})_{kj}} = e{\delta _{ij}},~ {b_{AX_i^ + X_j^ + }} = {b_{A\Psi _k^ + \Psi _k^ + }}{({\mathcal{C}_R})_{ki}}({\mathcal{C}_R})_{kj}^* = e{\delta _{ij}},
\nonumber\\
&& {a_{ZX_i^ + X_j^ + }} = {a_{Z\Psi _k^ + \Psi _k^ + }}({\mathcal{C}_L})_{ki}^*{({\mathcal{C}_L})_{kj}},~ {b_{ZX_i^ + X_j^ + }} = {b_{Z\Psi _k^ + \Psi _k^ + }}{({\mathcal{C}_R})_{ki}}({\mathcal{C}_R})_{kj}^*,
\nonumber\\
&& {a_{ZX_i^0X_j^0}} = {a_{Z\Psi _k^0\Psi _k^0}}\mathcal{N}_{ki}^*{\mathcal{N}_{kj}},~ {b_{ZX_i^0X_j^0}} = {b_{Z\Psi _k^0\Psi _k^0}}{\mathcal{N}_{ki}}\mathcal{N}_{kj}^*,
\nonumber\\
&& {a_{A{X^{ +  + }}{X^{ +  + }}}} = {b_{A{X^{ +  + }}{X^{ +  + }}}} = 2e,~ {a_{Z{X^{ +  + }}{X^{ +  + }}}} = {b_{Z{X^{ +  + }}{X^{ +  + }}}} = \frac{g}{{2{c_W}}}(3c_W^2 - s_W^2),
\nonumber\\
&& {a_{WX_i^ + X_j^0}} = {a_{W\Psi _k^ + \Psi _k^0}}({\mathcal{C}_L})_{ki}^*{\mathcal{N}_{kj}},~ {b_{WX_i^ + X_j^0}} = {b_{W\Psi _k^ + \Psi _k^0}}{({\mathcal{C}_R})_{ki}}\mathcal{N}_{kj}^*,
\nonumber\\
&& {a_{W{X^{ +  + }}X_i^ + }} = {a_{W{X^{ +  + }}\Psi _j^ + }}{({\mathcal{C}_L})_{ji}},~ {b_{W{X^{ +  + }}X_i^ + }} = {b_{W{X^{ +  + }}\Psi _j^ + }}({\mathcal{C}_R})_{ji}^*,
\nonumber\\
&& {a_{hX_i^0X_j^0}} = {a_{h\Psi _k^0\Psi _l^0}}{\mathcal{N}_{ki}}{\mathcal{N}_{lj}},~ {b_{hX_i^0X_j^0}} = {b_{h\Psi _k^0\Psi _l^0}}\mathcal{N}_{ki}^*\mathcal{N}_{lj}^*,
\nonumber\\
&& {a_{{G^0}X_i^0X_j^0}} = {a_{{G^0}\Psi _k^0\Psi _l^0}}{\mathcal{N}_{ki}}{\mathcal{N}_{lj}},~ {b_{{G^0}X_i^0X_j^0}} = {b_{{G^0}\Psi _k^0\Psi _l^0}}\mathcal{N}_{ki}^*\mathcal{N}_{lj}^*,
\nonumber\\
&& {a_{hX_i^ + X_j^ + }} = {a_{h\Psi _k^ + \Psi _l^ + }}{({\mathcal{C}_R})_{ki}}{({\mathcal{C}_L})_{lj}},~ {b_{hX_i^ + X_j^ + }} = {b_{h\Psi _k^ + \Psi _l^ + }}({\mathcal{C}_L})_{ki}^*({\mathcal{C}_R})_{lj}^*,
\nonumber\\
&& {a_{{G^0}X_i^ + X_j^ + }} = {a_{{G^0}\Psi _k^ + \Psi _l^ + }}{({\mathcal{C}_R})_{ki}}{({\mathcal{C}_L})_{lj}},~ {b_{{G^0}X_i^ + X_j^ + }} = {b_{{G^0}\Psi _k^ + \Psi _l^ + }}({\mathcal{C}_L})_{ki}^*({\mathcal{C}_R})_{lj}^*,
\nonumber\\
&& {a_{{G^ \pm }X_i^ + X_j^0}} = {a_{{G^ \pm }\Psi _k^ + \Psi _l^0}}{({\mathcal{C}_R})_{ki}}{\mathcal{N}_{lj}},~ {b_{{G^ \pm }X_i^ + X_j^0}} = {b_{{G^ \pm }\Psi _k^ + \Psi _l^0}}({\mathcal{C}_L})_{ki}^*\mathcal{N}_{lj}^*,
\nonumber\\
&& {a_{{G^ \pm }{X^{ +  + }}X_i^ + }} = {a_{{G^ \pm }{X^{ +  + }}\Psi _j^ + }}{({\mathcal{C}_L})_{ji}},~ {b_{{G^ \pm }{X^{ +  + }}X_i^ + }} = {b_{{G^ \pm }{X^{ +  + }}\Psi _j^ + }}({\mathcal{C}_R})_{ji}^* .
\end{eqnarray}

\section{Self Energies}
\label{app:self}

In this appendix, we give useful expressions for the self-energies of $\chi_i^0$, $\chi_i^\pm$, $\chi^{\pm\pm}$, $\gamma$, $Z$, and $W$,
used for both the calculation of the mass corrections for dark sector fermions and the electroweak oblique parameters.
In these calculations, we use the one-loop integrals whose definitions are consistent with Ref.~\cite{Denner:1991kt}:
\begin{eqnarray}
{A_0}({m^2}) &=& \frac{{{{(2\pi \mu )}^{4 - D}}}}{{i{\pi ^2}}}\int {{d^D}q} \frac{1}{{{q^2} - {m^2} + i\varepsilon }},
\\
{B_0}({p^2},m_1^2,m_2^2) &=& \frac{{{{(2\pi \mu )}^{4 - D}}}}{{i{\pi ^2}}}\int {{d^D}q} \frac{1}{{[{q^2} - m_1^2 + i\varepsilon ][{{(p + q)}^2} - m_2^2 + i\varepsilon ]}},
\\
{p_\mu }{B_1}({p^2},m_1^2,m_2^2) &=& \frac{{{{(2\pi \mu )}^{4 - D}}}}{{i{\pi ^2}}}\int {{d^D}q} \frac{{{q_\mu }}}{{[{q^2} - m_1^2 + i\varepsilon ][{{(p + q)}^2} - m_2^2 + i\varepsilon ]}},
\\
{g_{\mu \nu }}{B_{00}}({p^2},m_1^2,m_2^2) + {p_\mu }{p_\nu }{B_{11}}({p^2},m_1^2,m_2^2) \hspace*{-9em} &&
\nonumber\\
&=& \frac{{{{(2\pi \mu )}^{4 - D}}}}{{i{\pi ^2}}}\int {{d^D}q} \frac{{{q_\mu }{q_\nu }}}{{[{q^2} - m_1^2 + i\varepsilon ][{{(p + q)}^2} - m_2^2 + i\varepsilon ]}}.
\end{eqnarray}

We calculate the self-energies of $\chi_i^0$, $\chi_i^\pm$, and $\chi^{\pm\pm}$ in the $\overline{\mathrm{DR}}$ 
scheme with the 't~Hooft-Feynman gauge, as in Ref.~\cite{Pierce:1996zz}.
At NLO, the $\chi_i^0$-$\chi_k^0$ self-energy has contributions from loops 
of $W^\pm\chi_j^\mp$, $G^\pm\chi_j^\mp$, $h\chi_j^0$, $Z\chi_j^0$, and $G^0\chi_j^0$. We have
\begin{eqnarray}
&&16{\pi ^2}\Sigma _{X_i^0X_k^0}^{\mathrm{LV}}({p^2})
\nonumber\\
 &=& \sum\limits_j {( - 2a_{WX_j^ + X_i^0}^*{a_{WX_j^ + X_k^0}} - 2{b_{WX_j^ + X_i^0}}b_{WX_j^ + X_k^0}^* - a_{{G^ \pm }X_j^ + X_i^0}^*{a_{{G^ \pm }X_j^ + X_k^0}}} 
\nonumber\\
&&\qquad - {b_{{G^ \pm }X_j^ + X_i^0}}b_{{G^ \pm }X_j^ + X_k^0}^*){B_1}({p^2},m_{\chi _j^ \pm }^2,m_W^2) - \sum\limits_j {{b_{hX_i^0X_j^0}}{a_{hX_j^0X_k^0}}} {B_1}({p^2},m_{\chi _j^0}^2,m_h^2)
\nonumber\\
&& + \sum\limits_j {( - 2{a_{ZX_i^0X_j^0}}{a_{ZX_j^0X_k^0}} - {b_{{G^0}X_i^0X_j^0}}{a_{{G^0}X_j^0X_k^0}}){B_1}({p^2},m_{\chi _j^0}^2,m_Z^2)},
\\
&&16{\pi ^2}\Sigma _{X_i^0X_k^0}^{\mathrm{LS}}({p^2})
\nonumber\\
 &=& \sum\limits_j {( - 4b_{WX_j^ + X_i^0}^*{a_{WX_j^ + X_k^0}} - 4{a_{WX_j^ + X_i^0}}b_{WX_j^ + X_k^0}^* + b_{{G^ \pm }X_j^ + X_i^0}^*{a_{{G^ \pm }X_j^ + X_k^0}}} 
\nonumber\\
&&\quad + {a_{{G^ \pm }X_j^ + X_i^0}}b_{{G^ \pm }X_j^ + X_k^0}^*){m_{\chi _j^ \pm }}{B_0}({p^2},m_{\chi _j^ \pm }^2,m_W^2) + \sum\limits_j {{a_{hX_i^0X_j^0}}{a_{hX_j^0X_k^0}}} {m_{\chi _j^0}}{B_0}({p^2},m_{\chi _j^0}^2,m_h^2)
\nonumber\\
&& + \sum\limits_j {( - 4{b_{ZX_i^0X_j^0}}{a_{ZX_j^0X_k^0}} + {a_{{G^0}X_i^0X_j^0}}{a_{{G^0}X_j^0X_k^0}}){m_{\chi _j^0}}{B_0}({p^2},m_{\chi _j^0}^2,m_Z^2)}.
\end{eqnarray}
The $\chi_i^+$-$\chi_k^+$ self-energy has contributions from loops of $W^+\chi_j^0$, $G^+\chi_j^0$, $Z\chi_j^+$, $G^0\chi_j^+$, $A\chi_j^+$, $h\chi_j^+$, $W^-\chi^{++}$, and $G^-\chi^{++}$.  Therefore,
\begin{eqnarray}
&&16{\pi ^2}\Sigma _{X_i^ + X_k^ + }^{\mathrm{LV}}({p^2})
\nonumber\\
 &=& \sum\limits_j {( - 2{a_{WX_i^ + X_j^0}}a_{WX_k^ + X_j^0}^* - {b_{{G^ \pm }X_i^ + X_j^0}}b_{{G^ \pm }X_k^ + X_j^0}^*){B_1}({p^2},m_{\chi _j^0}^2,m_W^2)} 
\nonumber\\
&& + \sum\limits_j {( - 2{a_{ZX_i^ + X_j^ + }}{a_{ZX_j^ + X_k^ + }} - {b_{{G^0}X_i^ + X_j^ + }}{a_{{G^0}X_j^ + X_k^ + }}){B_1}({p^2},m_{\chi _j^ \pm }^2,m_Z^2)} 
\nonumber\\
&& - 2\sum\limits_j {{a_{AX_i^ + X_j^ + }}{a_{AX_j^ + X_k^ + }}} {B_1}({p^2},m_{\chi _j^ \pm }^2,0) - \sum\limits_j {{b_{hX_i^ + X_j^ + }}{a_{hX_j^ + X_k^ + }}} {B_1}({p^2},m_{\chi _j^ \pm }^2,m_h^2)
\nonumber\\
&& + ( - 2a_{W{X^{ +  + }}X_i^ + }^*{a_{W{X^{ +  + }}X_k^ + }} - a_{{G^ \pm }{X^{ +  + }}X_i^ + }^*{a_{{G^ \pm }{X^{ +  + }}X_k^ + }}){B_1}({p^2},m_{{\chi ^{ \pm  \pm }}}^2,m_W^2),
\\
&&16{\pi ^2}\Sigma _{X_i^ + X_k^ + }^{\mathrm{LS}}({p^2})
\nonumber\\
 &=& \sum\limits_j {( - 4{b_{WX_i^ + X_j^0}}a_{WX_k^ + X_j^0}^* + {a_{{G^ \pm }X_i^ + X_j^0}}b_{{G^ \pm }X_k^ + X_j^0}^*){m_{\chi _j^0}}{B_0}({p^2},m_{\chi _j^0}^2,m_W^2)} 
\nonumber\\
&& + \sum\limits_j {( - 4{b_{ZX_i^ + X_j^ + }}{a_{ZX_j^ + X_k^ + }} + {a_{{G^0}X_i^ + X_j^ + }}{a_{{G^0}X_j^ + X_k^ + }}){m_{\chi _j^ \pm }}{B_0}({p^2},m_{\chi _j^ \pm }^2,m_Z^2)} 
\nonumber\\
&& - 4\sum\limits_j {{b_{AX_i^ + X_j^ + }}{a_{AX_j^ + X_k^ + }}} {m_{\chi _j^ \pm }}{B_0}({p^2},m_{\chi _j^ \pm }^2,0) + \sum\limits_j {{a_{hX_i^ + X_j^ + }}{a_{hX_j^ + X_k^ + }}} {m_{\chi _j^ \pm }}{B_0}({p^2},m_{\chi _j^ \pm }^2,m_h^2)
\nonumber\\
&& + ( - 4b_{W{X^{ +  + }}X_i^ + }^*{a_{W{X^{ +  + }}X_k^ + }} + b_{{G^ \pm }{X^{ +  + }}X_i^ + }^*{a_{{G^ \pm }{X^{ +  + }}X_k^ + }}){m_{{\chi ^{ \pm  \pm }}}}{B_0}({p^2},m_{{\chi ^{ \pm  \pm }}}^2,m_W^2).
\end{eqnarray}
The self-energy of $\chi^{++}$ receives contributions from loops of $W^+\chi_i^+$, $G^+\chi_i^+$, $A\chi^{++}$, and $Z\chi^{++}$, and thus
\begin{eqnarray}
&&16{\pi ^2}\Sigma _{{X^{ +  + }}{X^{ +  + }}}^{\mathrm{LV}}({p^2})
\nonumber\\
 &=& \sum\limits_i {( - 2|{a_{W{X^{ +  + }}X_i^ + }}{|^2} - |{b_{{G^ \pm }{X^{ +  + }}X_i^ + }}{|^2}){B_1}({p^2},m_{\chi _i^ + }^2,m_W^2)} 
\nonumber\\
&& - 2a_{A{X^{ +  + }}{X^{ +  + }}}^2{B_1}({p^2},m_{{\chi ^{ \pm  \pm }}}^2,0) - 2a_{Z{X^{ +  + }}{X^{ +  + }}}^2{B_1}({p^2},m_{{\chi ^{ \pm  \pm }}}^2,m_Z^2),
\\
&&16{\pi ^2}\Sigma _{{X^{ +  + }}{X^{ +  + }}}^{\mathrm{LS}}({p^2})
\nonumber\\
 &=& \sum\limits_i {( - 4{b_{W{X^{ +  + }}X_i^ + }}a_{W{X^{ +  + }}X_i^ + }^* + {a_{{G^ \pm }{X^{ +  + }}X_i^ + }}b_{{G^ \pm }{X^{ +  + }}X_i^ + }^*){m_{\chi _i^ + }}{B_0}({p^2},m_{\chi _i^ + }^2,m_W^2)} 
\nonumber\\
&& - 4{b_{A{X^{ +  + }}{X^{ +  + }}}}{a_{A{X^{ +  + }}{X^{ +  + }}}}{m_{{\chi ^{ \pm  \pm }}}}{B_0}({p^2},m_{{\chi ^{ \pm  \pm }}}^2,0)
\nonumber\\
&& - 4{b_{Z{X^{ +  + }}{X^{ +  + }}}}{a_{Z{X^{ +  + }}{X^{ +  + }}}}{m_{{\chi ^{ \pm  \pm }}}}{B_0}({p^2},m_{{\chi ^{ \pm  \pm }}}^2,m_Z^2).
\end{eqnarray}
The expressions for ${\Sigma ^{{\mathrm{RV}}}}({p^2})$ and ${\Sigma ^{{\mathrm{RS}}}}({p^2})$ can be obtained from ${\Sigma ^{{\mathrm{LV}}}}({p^2})$ and ${\Sigma ^{{\mathrm{LS}}}}({p^2})$
through $a\leftrightarrow b$, respectively.

Below we give the extra contributions to the vacuum polarization amplitudes of electroweak gauge bosons from the triplet and quadruplets. The contribution to the $Z$ boson vacuum polarization comes from loops of $\chi_i^0 \chi_j^0$, $\chi_i^+ \chi_j^-$, and $\chi^{++}\chi^{--}$:
\begin{eqnarray}
16{\pi ^2}\Delta {\Pi _{ZZ}}({p^2}) &=& \frac{1}{2}\sum\limits_{ij} {[({a_{ZX_j^0X_i^0}}{a_{ZX_i^0X_j^0}} + {b_{ZX_j^0X_i^0}}{b_{ZX_i^0X_j^0}}){J_1}({p^2},m_{\chi _i^0}^2,m_{\chi _j^0}^2)} 
\nonumber\\
&&\quad - 2({a_{ZX_j^0X_i^0}}{b_{ZX_i^0X_j^0}} + {b_{ZX_j^0X_i^0}}{a_{ZX_i^0X_j^0}}){m_{\chi _i^0}}{m_{\chi _j^0}}{B_0}({p^2},m_{\chi _i^0}^2,m_{\chi _j^0}^2)]
\nonumber\\
&& + \sum\limits_{ij} {[({a_{ZX_j^ + X_i^ + }}{a_{ZX_i^ + X_j^ + }} + {b_{ZX_j^ + X_i^ + }}{b_{ZX_i^ + X_j^ + }}){J_1}({p^2},m_{\chi _i^ \pm }^2,m_{\chi _j^ \pm }^2)} 
\nonumber\\
&&\quad - 2({a_{ZX_j^ + X_i^ + }}{b_{ZX_i^ + X_j^ + }} + {b_{ZX_j^ + X_i^ + }}{a_{ZX_i^ + X_j^ + }}){m_{\chi _i^ \pm }}{m_{\chi _j^ \pm }}{B_0}({p^2},m_{\chi _i^ \pm }^2,m_{\chi _j^ \pm }^2)]
\nonumber\\
&& + \frac{{{g^2}{{(3c_W^2 - s_W^2)}^2}}}{{2c_W^2}}{J_2}({p^2},m_{{\chi ^{ \pm  \pm }}}^2),
\end{eqnarray}
where
\begin{eqnarray}
{J_1}({p^2},m_1^2,m_2^2) &\equiv& {A_0}(m_1^2) + {A_0}(m_2^2) - ({p^2} - m_1^2 - m_2^2){B_0}({p^2},m_1^2,m_2^2) - 4{B_{00}}({p^2},m_1^2,m_2^2),
\nonumber\\
{J_2}({p^2},{m^2}) &\equiv& 2{A_0}({m^2}) - {p^2}{B_0}({p^2},{m^2},{m^2}) - 4{B_{00}}({p^2},{m^2},{m^2}).
\end{eqnarray}
The contribution to the $W^+$ boson vacuum polarization comes from loops of $\chi_i^0 \chi_j^+$ and $\chi_i^- \chi^{++}$:
\begin{eqnarray}
&& 16{\pi ^2}\Delta {\Pi _{WW}}({p^2})
\nonumber\\
&=& \sum\limits_{ij} {[(|{a_{WX_j^ + X_i^0}}{|^2} + |{b_{WX_j^ + X_i^0}}{|^2}){J_1}({p^2},m_{\chi _i^0}^2,m_{\chi _j^ \pm }^2)} 
\nonumber\\
&&\quad - 2({a_{WX_j^ + X_i^0}}b_{WX_j^ + X_i^0}^* + {b_{WX_j^ + X_i^0}}a_{WX_j^ + X_i^0}^*){m_{\chi _i^0}}{m_{\chi _j^ \pm }}{B_0}({p^2},m_{\chi _i^0}^2,m_{\chi _j^ \pm }^2)]
\nonumber\\
&& + \sum\limits_i {[(|{a_{W{X^{ +  + }}X_i^ + }}{|^2} + |{b_{W{X^{ +  + }}X_i^ + }}{|^2}){J_1}({p^2},m_{\chi _i^ \pm }^2,m_{{\chi ^{ \pm  \pm }}}^2)} 
\nonumber\\
&&\quad - 2({a_{W{X^{ +  + }}X_i^ + }}b_{W{X^{ +  + }}X_i^ + }^* + {b_{W{X^{ +  + }}X_i^ + }}a_{W{X^{ +  + }}X_i^ + }^*){m_{\chi _i^ \pm }}{m_{{\chi ^{ \pm  \pm }}}}{B_0}({p^2},m_{\chi _i^ \pm }^2,m_{{\chi ^{ \pm  \pm }}}^2)].
\nonumber\\*
\end{eqnarray}
The contribution to the photon vacuum polarization comes from loops of $\chi_i^+ \chi_i^-$ and $\chi^{++}\chi^{--}$:
\begin{equation}
16{\pi ^2}\Delta {\Pi _{AA}}({p^2}) = 2{e^2}\sum\limits_i {{J_2}({p^2},m_{\chi _i^ \pm }^2)}  + 8{e^2}{J_2}({p^2},m_{{\chi ^{ \pm  \pm }}}^2).
\end{equation}
And finally, the contribution to the mixed photon-$Z$ vacuum polarization also arises from 
loops of $\chi_i^+ \chi_i^-$ and $\chi^{++}\chi^{--}$:
\begin{equation}
16{\pi ^2}\Delta {\Pi _{ZA}}({p^2}) = e\sum\limits_i {({a_{ZX_i^ + X_i^ + }} + {b_{ZX_i^ + X_i^ + }}){J_2}({p^2},m_{\chi _i^ \pm }^2)}  + \frac{{2eg(3c_W^2 - s_W^2)}}{{{c_W}}}{J_2}({p^2},m_{{\chi ^{ \pm  \pm }}}^2).
\end{equation}

\bibliographystyle{JHEP}
\bibliography{QTH}

\end{document}